\documentclass[12pt,authoryear]{aastex}



\sfcode`A=1000 \sfcode`B=1000 \sfcode`C=1000 \sfcode`D=1000
\sfcode`E=1000 \sfcode`F=1000 \sfcode`G=1000 \sfcode`H=1000
\sfcode`I=1000 \sfcode`J=1000 \sfcode`K=1000 \sfcode`L=1000
\sfcode`M=1000 \sfcode`N=1000 \sfcode`O=1000 \sfcode`P=1000
\sfcode`Q=1000 \sfcode`R=1000 \sfcode`S=1000 \sfcode`T=1000
\sfcode`U=1000 \sfcode`V=1000 \sfcode`W=1000 \sfcode`X=1000
\sfcode`Y=1000 \sfcode`Z=1000

\newcommand{\Msun}{\mbox{M$_\sun$}}

\newcommand{\sst}{{\it Spitzer Space Telescope}}
\newcommand{\s}{{\it Spitzer}}
\newcommand{\h}{{\it Herschel}}

\newcommand{\mmjy}{\hbox{$\mu$Jy}}

\slugcomment{\today}
\shorttitle{The K-band Luminosity Function}
\begin{document}
\title{Multi-Wavelength Study of a Complete IRAC
  3.6\,\micron-Selected Galaxy Sample: a Fair Census of Red and Blue
  Populations at Redshifts 0.4--1.2} 
\author{J.-S.~Huang,$\!$\altaffilmark{1,2}
S. M.~Faber,$\!$\altaffilmark{3}
C. N. A.~Willmer,$\!$\altaffilmark{4}
D.~Rigopoulou,$\!$\altaffilmark{5}
D.~Koo,$\!$\altaffilmark{3}
J.~Newman,$\!$\altaffilmark{6}
C.~Shu,$\!$\altaffilmark{7}
M. L. N. Ashby,$\!$\altaffilmark{2}
P.~Barmby,$\!$\altaffilmark{8}
A.~Coil,$\!$\altaffilmark{9}
Z.~Luo,$\!$\altaffilmark{7}
G.~Magdis,$\!$\altaffilmark{5}
T.~Wang,$\!$\altaffilmark{10,2}
B.~Weiner,$\!$\altaffilmark{4}
S.~P.~Willner,$\!$\altaffilmark{2}
X. Z.~Zheng,$\!$\altaffilmark{11}
\& G.~G.~Fazio$\!$\altaffilmark{2}
}
\altaffiltext{1}{National Astronomical Observatories, Chinese Academy of Sciences, Beijing 100012, China}
\altaffiltext{2}{Harvard-Smithsonian Center for Astrophysics, 60 Garden Street, MS65, Cambridge, MA 02138, USA}
\altaffiltext{3}{University of California Observatories/Lick Observatory, University of California, Santa Cruz, CA 95064, USA}
\altaffiltext{4}{Steward Observatory, University of Arizona, 933 North Cherry Avenue, Tucson, AZ 85721, USA}
\altaffiltext{5}{Department of Astrophysics, Oxford University, Keble Road, Oxford, OX1 3RH, UK}
\altaffiltext{6}{Department of Physics and Astronomy, University of Pittsburgh, 3941 O'Hara Street, Pittsburgh, PA 15260, USA}
\altaffiltext{7}{Shanghai Key Lab for Astrophysics, Shanghai Normal University, 100 Guilin Road, Shanghai 200234, P. R. China}
\altaffiltext{8}{Department of Physics and Astronomy, University of Western Ontario, 1151 Richmond Street, London, ON N6A 3K7, Canada}
\altaffiltext{9}{Department of Physics and Center for Astrophysics and Space Sciences, University of California, San Diego, 9500 Gilman Drive, La Jolla, CA 92093, USA}
\altaffiltext{10}{School of Astronomy and Space Science, Nanjing University, Nanjing, China}
\altaffiltext{11}{Purple Mountain Observatory, Nanjing, P. R. China}

\begin{abstract}
 
  We present a multi-wavelength study of a 3.6\,$\mu$m-selected
  galaxy sample in the Extended Groth strip. The sample is complete
  for galaxies with stellar mass $>$10$^{9.5}$\,\Msun and redshift
  $0.4<z<1.2$. In this redshift range, the IRAC 3.6\,$\mu$m band
  measures the rest-frame near-infrared band, permitting nearly
  unbiased selection with respect to both quiescent and star-forming
  galaxies. The numerous spectroscopic redshifts available in the EGS
  are used to train an Artificial Neural Network to estimate
  photometric redshifts. The distribution of photometric redshift
  errors is Gaussian with standard deviation ${\sim}0.025(1+z)$, and
  the fraction of redshift failures (${>}3\sigma$ errors) is about
  3.5\%. A new method of validation based on pair statistics confirms
  the estimate of standard deviation even for galaxies lacking
  spectroscopic redshifts.  Basic galaxy properties measured include
  rest-frame $U-B$ colors, $B$- and $K$-band absolute magnitudes, and
  stellar masses. We divide the sample into quiescent and
  star-forming galaxies according to their rest-frame $U-B$ colors
  and 24 to 3.6\,\micron\ flux density ratios and derive rest
  $K$-band luminosity functions and stellar mass functions for
  quiescent, star forming, and all galaxies.  The results show that
  massive, quiescent galaxies were in place by $z\approx1$, but lower
  mass galaxies generally ceased their star formation at later epochs.

\end{abstract}



\section{INTRODUCTION}

The central puzzle of galaxy evolution is how baryons fall into
galaxies and turn into stars. Understanding the mass assembly history
of galaxies is a critical element for solving this puzzle. Galaxies
can be generally divided into two populations: those 
actively forming stars and those that are quiescent with relatively
little star formation. Observationally, these two types of galaxy
have differing colors corresponding to different regions in color-magnitude diagrams \citep{str01,
  hog03}.  The star-forming galaxies are said to occupy the ``blue
cloud'' (BC) while the passive galaxies occupy the``red
sequence'' (RS).  Similar bimodal distributions are also seen in their
morphologies, spectral types, metallicities, and star formation rates
\citep{Madgwick2003,kauffmann2003a,kauffmann2003b}. Morphologically
most galaxies in the RS are elliptical/S0 galaxies, and
galaxies in the BC are disk or irregular/peculiar
galaxies. This bimodality is observed over a very wide redshift
range up to $z\la2.5$ \citep{bel04,willmer2006,faber2007,brammer2009}.
In a color--stellar mass diagram, red galaxies appear to be much more
massive than blue ones. 

Formation of massive red galaxies cannot be accounted for by any
simple blue-to-red galaxy evolution path.  Many authors \citep{tt72,
  toth92, kau93, col96, hopkins2006} suggest that galaxy merging is
the only way of forming massive red galaxies.  There are probably
more than two merging scenarios: two blue galaxies can merge to
trigger a burst to star formation and eventually form a massive red
galaxy, or two blue galaxies can evolve to red galaxies, then undergo
a dry merging to form a massive galaxy. A census of red and blue
galaxies at different redshifts is needed to constrain how galaxies
migrate from the BC to the RS.

Redshifts are essential for studying galaxy populations. Early
spectroscopic galaxy surveys \citep{lil95, cow96,lin1999,cohen2002}
usually obtained several hundreds of redshifts. These surveys
concluded that strong galaxy evolution seen in faint $B$-band galaxy
counts was mainly due to blue galaxy evolution since $z=1$ with red
galaxies showing little or no evolution. \citet{im2002} constructed
luminosity functions for elliptical galaxies in $0.5<z<1$ with the
DEEP1 redshift sample and argued that they saw only passive
luminosity evolution but no change in number density for elliptical
galaxies since $z=1$.  These early redshift surveys are too small to
have significant statistics for massive red galaxies, which can also
suffer a large cosmic variance \citep{som05}. Recently several large
optical redshift surveys have been carried out on 8\,m class
telescopes including VVDS \citep{lefevre2005}, $z$COSMOS
\citep{lilly2007}, and DEEP2 \citep{faber2007,davis07}, each with
more than 10000 redshifts available for galaxies at $0<z<1.5$. These
redshift surveys are highly complete for blue galaxies up to $z=1.4$
because the [\ion{O}{2}] $\lambda$3727 line can be
observed. Measuring redshifts for red galaxies is more difficult
because there are only absorption lines available in most of their
spectra, and at high redshifts, only luminous red galaxies can have
redshifts measured in these existing redshift surveys. Therefore
optical redshift samples suffer severe bias against red galaxies at
high redshifts.

Most studies of evolution of galaxy populations have used
optically-selected samples. Several groups \citep{bell2006,faber2007}
used photometric or photometric/spectroscopic combined redshift
samples to study evolution of red galaxies and found that the number
density of red galaxies increases from $z=1$ to $z=0$.  This is
consistent with a migration of galaxy population from BC to
RS.  However, as mentioned above, optically-selected
samples are biased against red galaxies at $z\sim 1$, where even $I$
band probes the rest-frame UV.

Near-IR selected galaxy samples permit detection of red galaxies with
lower luminosities or at higher redshifts \citep{brammer2009, pvd09}
than optically-selected samples.  The rest-frame $K$-band light of a
galaxy comes mainly from M and K stars and traces underlying stellar
mass\citep{cow96, hua97,Bell2001}. Therefore a rest-frame $K$-band
selected galaxy sample is equivalent to a mass-selected sample
unbiased against either red or blue galaxies, permitting fair
statistics for both galaxy populations. However, early ground-based
near infrared (NIR) surveys have been ineffective because of small format IR array
camera and high NIR sky background.  Large-format NIR array cameras
on 4--8\,m class telescopes now allow deep NIR surveys with both wide
coverage and faint limiting magnitude $K_{AB}\approx23$--25\,mag
\citep{cim02,franx2003,bundy2006,taylor2009, pvd09,mccracken2010}.
Most deep $K$-band surveys focus on formation and evolution of
passive red galaxies with $M_* >10^{11}$\,M$_{\odot}$ at $z\sim2$
\citep{franx2003,taylor2009,brammer2009, pvd09,mccracken2010}.  On
the other hand, local $K$-band galaxy surveys detect early type
galaxies with stellar masses as low as 10$^{9.5}$ M$_{\odot}$
\citep{cole2001, bell2003}. More recently, \citet{baldry2012} estimated the local stellar mass function down to M$_*$=10$^{7}$M$_{\odot}$, and
found that the local stellar mass function has a very steep slope at the lower mass end.
Most relevant to the present study,
\citet{bundy2006} carried out $K$-band imaging in the Extended Groth
Strip region with $K_{AB}<23$. The limiting stellar mass for their
$K$-selected sample is $\sim$10$^{10.5}$\,M$_{\odot}$ at
$z\sim1$. Formation and evolution of lower mass red galaxies since
$z=1$ has not yet been well studied.

The InfraRed Array Camera (IRAC) on board the \sst\ can perform very
fast scanning of the IR sky with great depth. The IRAC 3.6 and
4.5\,$\mu$m bands probe the rest NIR for galaxies at $z\sim 1$. There
are several large, deep IRAC surveys. Samples obtained in these
surveys are usually confusion limited at
$f_\nu\approx1$\,$\mu$Jy. The limiting flux of $1\mu$Jy
($[3.6]=23.9$\footnote{In this paper, the notation $[w]$ means the AB
  magnitude at wavelength $w$ in \micron.  Some other papers have
  used this notation to mean Vega magnitudes.}) corresponds to an
absolute $K$-band magnitude of $M_K=-19$ for a galaxy at $z=1$.  A
galaxy with $[3.6]=23.9$\,mag has a typical $R$-band magnitude of 26
\citep{barro2011}, too faint for spectroscopy even with 8\,m class
telescopes. Photometric redshifts are essential in studying faint
mid-infrared (MIR) selected galaxies.

Using photometric redshifts is inevitable in studying red galaxy
evolution.  With large format CCD cameras available on 6--8\,m class
telescopes, deep multi-wavelength photometric data become
available. These data permit measurement of accurate photometric
redshifts for much larger samples of faint galaxies than can be
obtained spectroscopically. A difficulty is that for galaxies at
higher redshifts, the dominant features in galaxy SEDs, for example
the Balmer/4000\AA\ break, shift out of the visible band.  Including
intermediate-band photometry can improve the photometric redshifts
\citep{wol03,ilbert2009}, and accuracy can be further improved by
extending the wavelength coverage to the NIR.  Recently the NEWFIRM
intermediate band survey, including 3 bands in the 1.2\,$\mu$m window
and 2 bands in the 1.6\,$\mu$m window, was completed with the NOAO
4\,m telescope in search of passive galaxies at $z>1.5$
\citep{whitaker2011}. The relative redshift uncertainty
$\sigma_z/(1+z)$ for their photometric redshifts is 1--2\% for the
$K$-selected galaxies.

To extend SED fitting to the \s\ IRAC bands, photometric
redshift estimation must overcome a new challenge: variation of point
spread function (PSF) from ultraviolet to the
mid-infrared. Traditionally, photometric redshift estimation
\citep{rudnick2001, brodwin2006, ilbert2006, ilbert2009} is performed
by fitting a set of templates to an observed galaxy spectral energy
distribution (SED) obtained in a multi-wavelength photometric survey.
PSF variation in different bands may distort the observed SEDs \citep{labbe2005}, and
anyway we have poor knowledge of galaxy SED templates in the MIR
\citep{lu2003, brodwin2006}. Artificial neural networks (ANN$z$,
Collister \& Lahav 2004) is an alternative redshift estimation
method. ANN$z$ has several advantages for MIR-selected galaxy
samples: it does not require any knowledge of galaxy SED templates,
and the resulting photometric redshifts are not affected by
photometry offsets caused by PSF variation or any other systematic
offset in different bands. ANN$z$ uses a calibration sample to set up
an empirical relation between spectroscopic redshifts and observed
galaxy properties. Input parameters can be anything including color,
flux, size, and even digital morphological parameters. All this
method requires is that input and calibration samples have the same
filter set for photometry and cover same input parameters and
redshift ranges. Though ANN$z$ cannot measure any redshifts beyond
the range of the calibration sample, \citet{brodwin2006} showed that
ANN$z$ photometric redshifts for a MIR-selected sample have much
smaller scatter than those derived with the SED template methods.

This paper presents a multi-wavelength study of a complete
3.6\,$\mu$m-selected galaxy sample at $z<1.2$ in the Extended Groth
Strip region. Photometric redshifts for this sample are derived using
ANN$z$. \S2 describes the sample selection, and \S3 explains the
photometric redshift derivation with ANN$z$ including the
uncertainties. \S4 gives results for the blue and red galaxy
populations at $0.4<z<1.2$ and their properties including the rest
$K$-band luminosity function and color-magnitude relation. \S5
summarizes results.  Source distances are based on standard flat
$\Lambda$CDM cosmology with $H_0=70$~km~s$^{-1}$~Mpc and $\Omega_M =
0.30$.

\section{The 3.6\,$\mu$m Selected Sample in EGS}

The All-wavelength Extended Groth Strip International Survey
\citep[AEGIS,][]{davis07} is a large multi-wavelength survey in bands
from X-ray \citep{laird2009} to 20\,cm \citep{ivison2007}.  
The original Chandra survey covers the whole strip (2$^{\circ} \times$ 15\arcmin) with 200ks per pointing, and
later the central 45\arcmin$\times$15\arcmin strip with a deeper exposure of 800ks per pointing. The x-ray
limiting flux densities \footnote{The limiting flux density is defined as the flux to which at least 1\% of the survey area is sensitive\citep{laird2009}.}  for the 200ks imaging in the whole strip are 5.3$\times$10$^{-17}$erg cm$^{-1}$ s$^{-1}$ in the soft (0.5-2 keV) band and 3.8$\times$10$^{-16}$erg cm$^{-1}$ s$^{-1}$ in the hard (2-10 keV) band\citep{laird2009}.
\citet{ivison2007} performed a deep radio imaging at 1.4GHz in the EGS region with the 5$\sigma$ limiting flux density of 50$\mu$Jy. Both X-ray and radio can help to identify
AGNs.
More recent data include {\it Herschel Space Telescope} FIR/submm imaging
in the EGS as part of the HerMES survey \citep{oliver2010,Roseboom2012,smith2012}. The SPIRE 5$\sigma$ limiting flux densities are 13.8, 11.3, and 16.4 mJy\citep{oliver2012}. The
angular resolution for the SPIRE images are rather poor, 18\arcsec, 25\arcsec, and 36\arcsec at 250, 350, and 500$\mu$m, sources are blended in the EGS SPIRE images. Deep 24\,\micron\ imaging is available in the EGS
region \citep{dickinson2006} with the 5$\sigma$ limiting flux density of
$f_{24}=20$\,$\mu$Jy, providing a good prior to identify the blended Herschel SPIRE sources. The
actual FIR photometry catalog in the ACS covered area was extracted using the MIPS 24$\mu$m catalog as the prior with the method described in \citet{Roseboom2012}.   \citet{rigopoulou2012} derived FIR luminosities for Herschel sources in the region by fitting templates to their IR SEDs.

The EGS also has
a large number of spectroscopic redshifts available
\citep{Davis2003,faber2007,Cooper2011,Cooper2012}.\footnote{http://deep.berkeley.edu/DR3/}. The
AEGIS datasets include one of the deepest IRAC surveys, covering
10\arcmin$\times$2\arcdeg\ with $\ga$9\,ks exposure time.  The IRAC
observations, data reduction, and photometry were presented by
\citet{barmby2008} and later with more aggressive source extraction
by \citet{Zheng2012}.  The latter authors found 80\% completeness at
limiting flux densities of 1.8 and 2.0\,$\mu$Jy respectively at
3.6\,\micron\ and 4.5\,\micron\ compared to $\sim$60\% completeness
at these flux densities found by \citeauthor{barmby2008}.  For the
current paper, we have restricted our sample to the HST ACS imaging
region (10\arcmin$\times$1\arcdeg, \citealt{Lotz2008}), where there
are generally deeper visible and NIR photometric data available.  Our
preliminary sample includes all objects in this region with flux
density $f_{3.6} > 2$\,$\mu$Jy in the \citeauthor{Zheng2012}
catalog. This limit selects galaxies over a wide redshift range and
also a few stars.  There are 23327 objects in the preliminary sample.

\begin{figure} \plotone{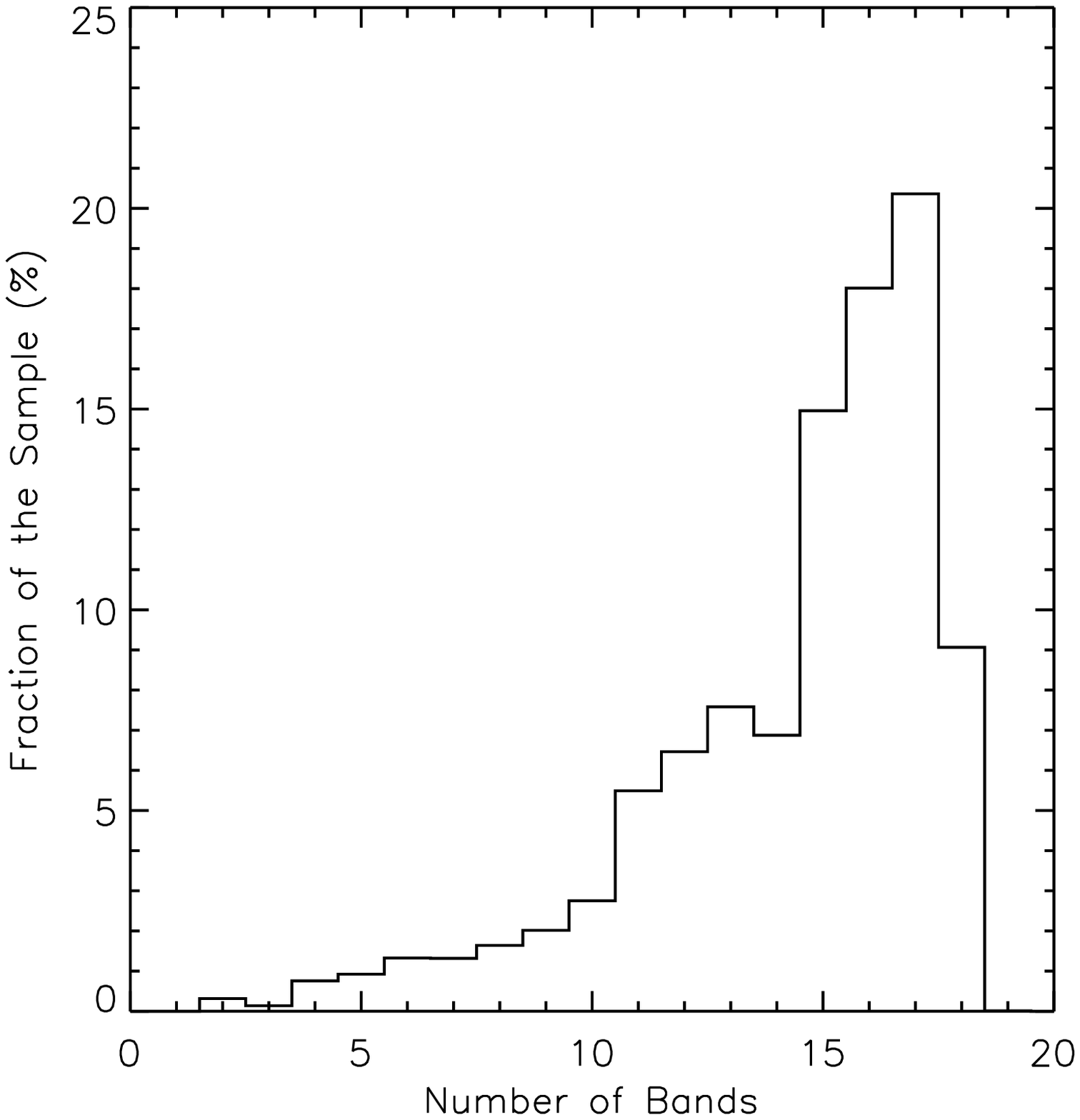}
  \caption{Histogram of number of detections in the multi-wavelength
    data set. There are a total of 18 photometric bands available for
    the sample.}
\label{f:nb}
\end{figure}

The visible and NIR photometric data for the sample are summarized in
Table~1. There are 18 bands from $u$ to 8\,\micron\ that are useful
for photometric redshifts. Figure~\ref{f:nb} shows a histogram of the
number of photometric points available per object. About 96\% of the
objects have photometry in at least 5 bands, and 62\% of the objects
have at least 15-band data.  As Table~1 shows, however, some of the
visible and NIR bands are redundant. For example, there are four
versions of $R$-band imaging with various depths.
In redundant bands, we used a $\sigma$-weighted mean to combine
photometry to give a single photometric value with maximum
signal-to-noise ratio in each band.  After combining redundant bands,
there are 12 wavelengths available in the $u$ to 8\,\micron\ range. In
the visible range {$ugriz$}, the Canada France Hawaii Telescope
Legacy Survey \citep[CFHTLS]{ilbert2006} provides a well-calibrated
photometry system, and we transformed all photometry in this
wavelength range into the CFTHLS system.

\begin{figure}
\epsscale{0.8}
\plotone{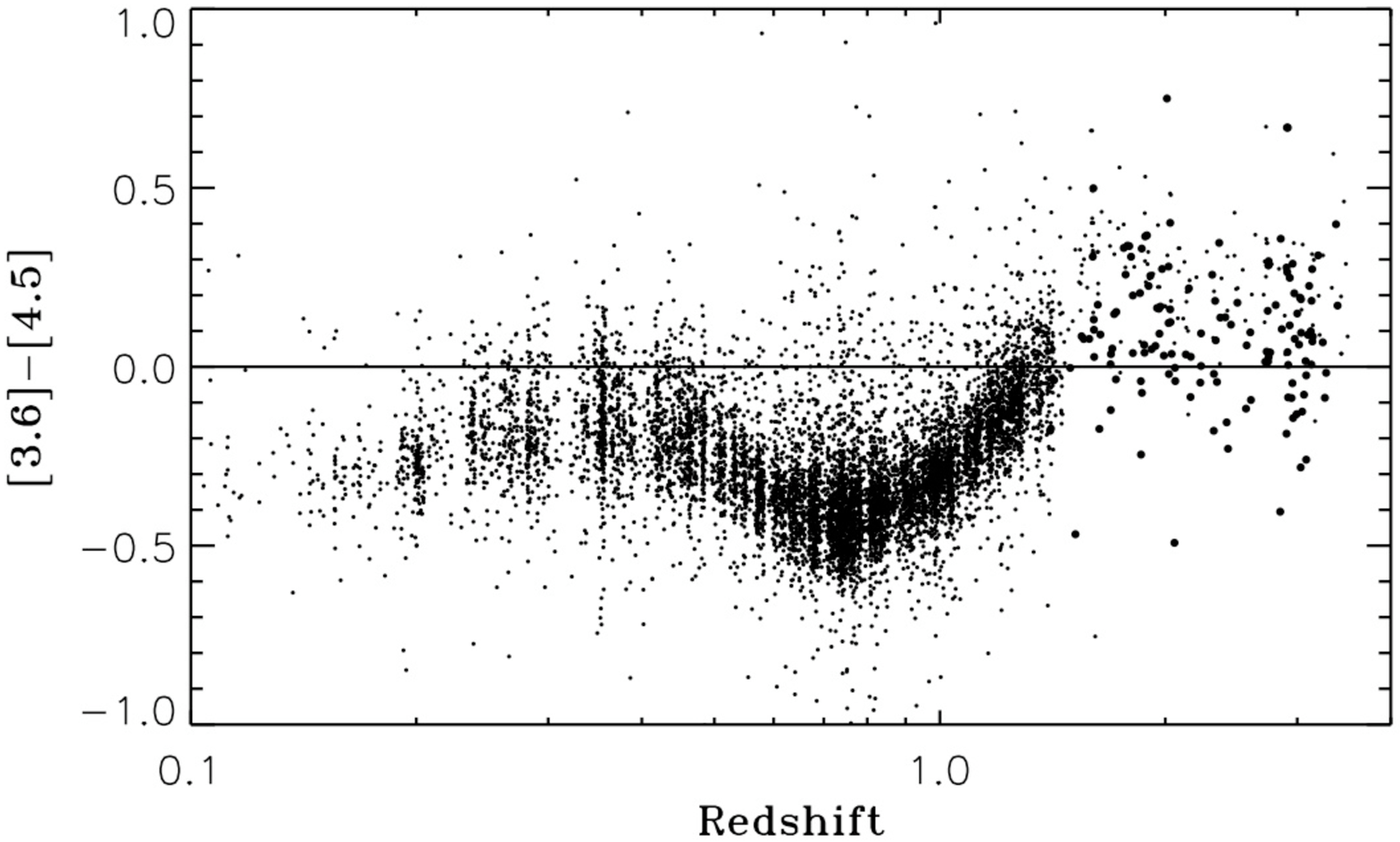}
\caption{IRAC $[3.6]-[4.5]$ colors as a function of redshift. All
  galaxies in the preliminary sample, i.e., before the color cuts of
  relations 1--3, are shown.  The fine dots denote the DEEP2
  galaxies, and bigger dots denote BM/BX galaxies and LBGs detected by IRAC
  \citep{steidel2003, steidel2004,shapley2005,hua05,rigopoulou2006} and
  MIPS 24\,$\mu$m selected ULIRGs at $z\sim 2$ \citep{huang2009}. The
  line at $[3.6]-[4.5]=0$ separates galaxies at $z\approx1.5$:
  galaxies with smaller redshift have bluer $[3.6]-[4.5]$ color.}
\label{f:cz}
\end{figure}

\begin{figure}
\epsscale{0.8}
\plotone{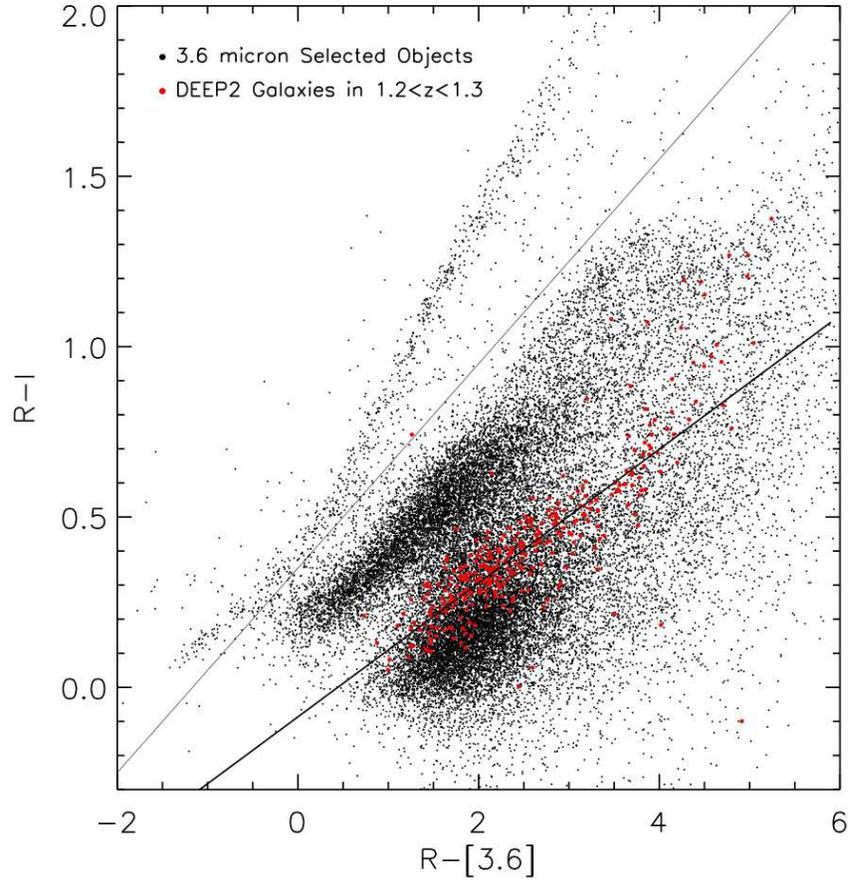}
\caption{Visible-IR color-color  diagram.  Points denote all objects
  selected in the 3.6\,\micron\
  preliminary sample; red points denote galaxies with spectroscopic
  redshifts between 1.20 and 1.30.  Lines show the color cuts of
  relations 2 (lower line) and 3 (upper line).  The final sample
  contains only objects in the middle zone of this diagram.}
\label{f:ri_ch1}
\end{figure}

This study focuses on galaxy population at $0.4<z<1.2$ to take
advantage of the large number of spectroscopic redshifts
available. Therefore the preliminary all-$z$ sample needs to be refined by
rejecting foreground stars and background galaxies at $z>1.2$. A set
of simple color criteria can separate the background
galaxies. \citet{hua04} and \citet{sorba2010} show that the IRAC
$[3.6]-[4.5]$ color is a function of redshift (Figure~\ref{f:cz}):
galaxies with $[3.6]-[4.5]<0$ are at $z<1.25,$ and galaxies with
$[3.6]-[4.5]>0$ are at $z>1.25$.\footnote{The exact redshift for
  galaxies with $[3.6]-[4.5]=0$ is weakly dependent on galaxy
  spectral types with red galaxies having $z=1.25$ and blue galaxies
  having $z=1.50$. \citet{sorba2010} proposed a more stringent color-dependent criterion to separate galaxies at z=1.3
: [3.6]-[4.5]=0.12([3.6]-[8.0])-0.07} This is because 1.6\,\micron\ bump in galaxy SEDs
shifts through the IRAC bands at $z>1$. Several groups have used this
SED feature to select galaxies at $z\approx2$ for \s\ IRS
spectroscopy with a nearly 100\% successful identification rate
\citep{weedman2006, Farrah2007, huang2009, desai2009,
  fadda2010}. Some rest-frame UV-selected galaxies at $2\la z\la3$ such
as BM/BX galaxies and LBGs, however, are too young to have developed the
1.6\,\micron\ bump \citep{hua05, rigopoulou2006}, and these galaxies
may have 
$[3.6]-[4.5]<0$ even  with $z>1.2$. We therefore need an additional criterion to reject
these young blue galaxies. We used an optical-MIR
color-color diagram ($R-I$ vs $R-[3.6]$, Figure~\ref{f:ri_ch1}) for
classification.  Figure~\ref{f:ri_ch1} shows that stars, galaxies,
at $z<1.25$, and galaxies at $z>1.25$ occupy different regions. Thus the
three criteria for selection of the 3.6\,\micron\ galaxy sample are:
\begin{equation} [3.6]-[4.6]<0
\end{equation} 
\begin{equation} R-I>0.2(R-[3.6])+0.015
\end{equation} 
\begin{equation} R-I<0.4(R-[3.6])+0.5
\end{equation}
Criterion~1 removes background galaxies with a normal 1.6\,\micron\
bump, criterion~2 rejects young, blue galaxies at $z>1.25$, and
criterion~3 rejects stars.  Criterion~1 is not as stringent as that proposed
by \citet{sorba2010} for the same purpose, thus may reduce the bias by the color cut, but will
include  galaxies at z$\ge$1.3 
 in the preliminary sample. These galaxies will be rejected from the sample after we apply Criterion~2.
Criterion~2 is very similar to the BzK selection for galaxies at z$\sim$2, but used here
to reject these galaxies. \citet{ilbert2009} used their SED-fiting photometric redshift sample to demonstrate that
galaxies can be well separated at z=1.3 in the same R$-$I vs R$-$[3.6] diagram.

\begin{figure}
\plotone{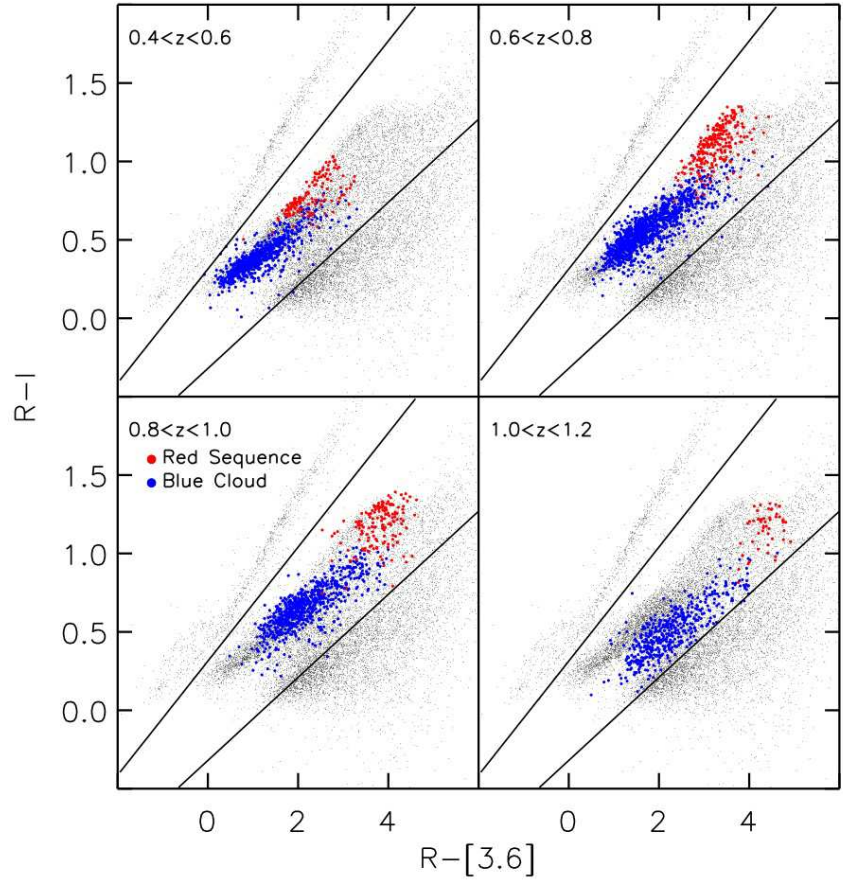}
\caption{Visible-IR color-color diagram for objects with redshifts and
  spectral classifications.  The locus of the preliminary sample is
  indicated in gray.  Red and blue points indicate galaxies in
  respectively the red sequence and blue cloud in the redshift bins
  indicated in each panel.}
\label{f:rich1_z}
\end{figure}

By combining both Criterion~1 and 2, we can effectively reject galaxies at z$>$1.25 while
reduce the bias caused by applying only one color criterion. 
The final sample for photometric redshift
estimation includes objects in the HST/ACS area of the EGS meeting the
above three criteria and with $[3.6]<23.15$. Figure~\ref{f:rich1_z}
confirms that our sample selection does not have any strong bias against either quiescent or star forming galaxies.  
There are 12890 galaxies in the final sample.

\begin{figure}
\epsscale{0.8}
\plotone{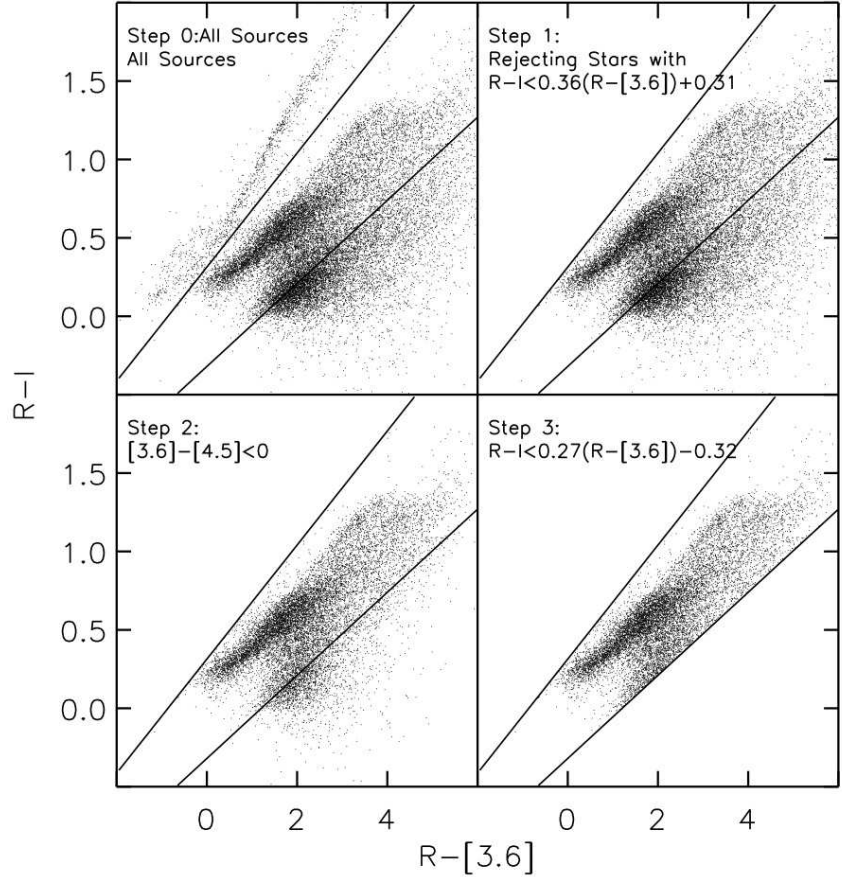}
\caption{Montage of visible-IR color-color diagrams as  sample
  selection criteria are successively applied.  The upper left panel
  shows all sources in the preliminary sample.  The panel to its
  right shows removal of stars via relation 3.  The lower left panel
  shows removal of most $z>1.25$ galaxies by relation 1, and the
  final panel shows removal of the rest by relation 2.}
\label{f:rich1_seq}
\end{figure}

Figure~\ref{f:rich1_seq} illustrates successive application of the
color criteria. In our preliminary sample, 95\% of the objects have
both $R$ and $I$ band detections and can be selected by all three
criteria. Of 1297 sources missing either $R$ or 
$I$, 1185 have  4.5\,\micron\ counterparts which we will be able to classify with relation~1: 
387 with [3.6]$-$[4.5]$<$0 are included in the sample and 798 with [3.6]$-$[4.5]$>$0 are reject.
The remaining 112 sources  cannot be definitively
excluded  and are kept in the sample, but their photometric redshifts
will be more uncertain than sources having better wavelength coverage.

Our selection does not particularly discriminate against AGNs, but AGNs with power-law SEDs in the mid-infrared band. 
Objects with strong hot dust emission are called "power-law" objects with very much red color of [3.6]$-$[4.5]$>$0 even at z$<$1 \citep{stern2005}. Thus
their 3.6$\mu$m flux densities for these power-law objects are mainly from
hot dust emission, but no longer trace their stellar mass. \citet{hua07} found that number of  power-law objects is less than 1\% of $>$L$_*$ galaxies in an 8\micron\ selected sample,
The 3.6\micron\ band is less sensitive to power-law objects than the 8\micron\ band, thus their fraction is even lower in the 3.6\micron\ selected sample. 
These objects will be rejected from the sample by Criterion 1.  

\section{Photometric Redshift Estimation}

\begin{figure}
\begin{flushleft}
\plotone{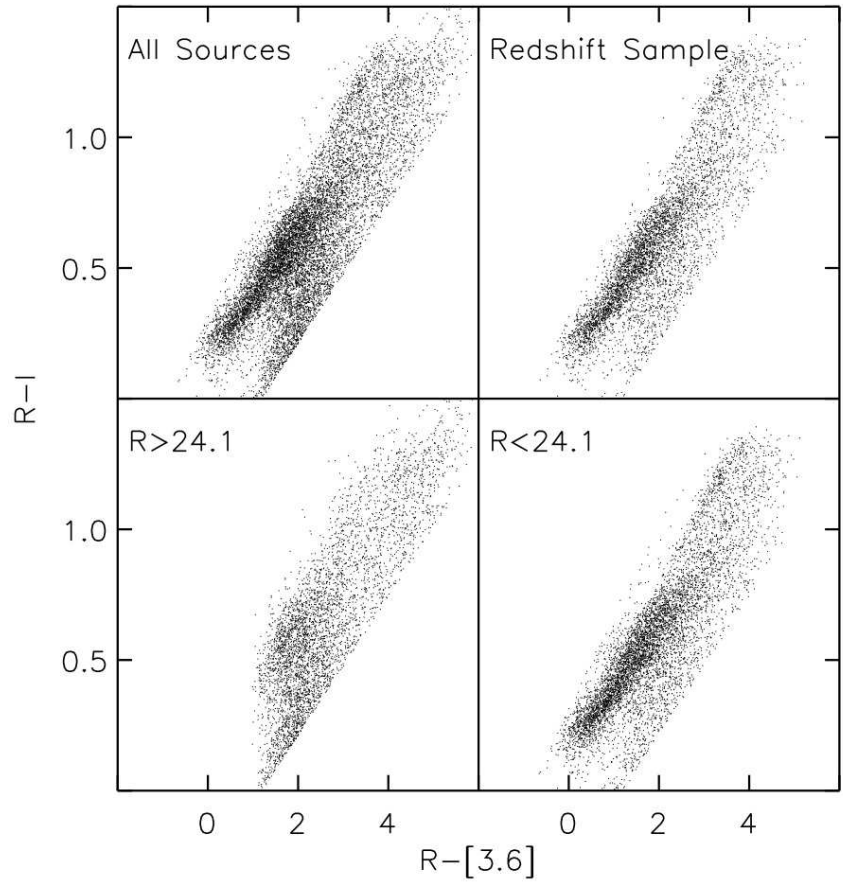}
\end{flushleft}
\caption{The color-color diagrams for the whole 3.6$\mu$m selected sample, the spectroscopic, faint (R$>$24.1) and bright (R$<$24.1)
subsamples. The limiting magnitude for the DEEP2+3 spectroscopic survey is R=24.1. This diagram shows that faint objects with
R$>$24.1 in the sample covers the same color range as the objects with spectroscopic redshifts, and we are able to estimate photometric redshifts for these optically faint objects with the spectroscopic subsample as the training set in ANNz.}
\label{f:faint_cc}
\end{figure}

The ANN$z$ method uses a subsample with known redshifts as a training
set to build up an empirical relation between measurable parameters
and redshifts and then applies this relation to the rest of the
sample to derive their photometric redshifts
\citep{firth2003,brodwin2006}. Almost any measurable parameter of
galaxies including magnitude, surface brightness, color, size, and
morphology can be used as an input node for ANN$z$, but magnitudes at
a range of wavelengths are the most common.  ANN$z$ requires that the
training set should cover the same ranges as the photometric sample
in all input parameters. In other words, ANN$z$ cannot derive
photometric redshift for a galaxy with input parameter values beyond
the range of its training set. The spectroscopic redshift sample has
a limiting magnitude of $R<24.1$, but a significant number of objects
in our sample are fainter than this limit. Using magnitudes as input
parameters to train the ANN$z$ will allow it to derive photometric
redshifts only for galaxies with $R<24.1$.  We therefore used colors
with 3.6\,$\mu$m photometry as the color base as the input nodes in
the ANN$z$ estimation.  There are 11 non-redundant colors for each
galaxy as shown in Table~1.  Faint galaxies ($R>24.1$) at $z<1.25$
exhibit the same color range as the galaxies with spectroscopic
redshifts shown in Figure~\ref{f:faint_cc}, and therefore the program will be able to drive
photometric redshifts for the whole sample.

ANN$z$ network architecture can be described as $N_{\rm
  input}:N_1:N_2:......:N_m:N_{\rm output}$, where $N_{\rm input}$ is
the number of input colors for each object. For the present data,
$N_{\rm input}=11$, the number of non-redundant colors. $N_1$, $N_2$,
.... and $N_m$ are the number of nodes in each ANN$z$ hidden layer
for a total of $m$ layers, and $N_{\rm output}$ is the number of sets
for output photometric redshift, usually set to be
1. \citet{firth2003} did extensive experimenting with ANN$z$ to
determine the optimal number of hidden layers for the best
photometric redshifts and found that a multi hidden layer
architecture yields photometric redshifts with a lower
$\sigma(z_p-z_s)$ than a single hidden layer architecture. A large
number of layers, however, do not significantly improve photometric
redshifts. \citet{firth2003} adopted 6:6:6:6:1 as their fiducial
architecture. We ran the ANN$z$ with various architectures for our
sample and arrived at the same conclusion. The final redshifts use
11:8:8:8:1 architecture, meaning 11 input colors, 3 hidden layers
with 8 nodes in each layer, and 1 output.

The accuracy of the ANN$z$ photometric redshifts is partly
determined by the number of objects in the training set.
\citet{firth2003} used a sample of 20000 objects to show that a
larger training set can yield more accurate photometric redshifts,
but increasing training set size beyond $n_{\rm train}=1000$ did not
improve the accuracy significantly. There are $\sim$5400
spectroscopic redshifts available in the sample. We divided the
spectroscopic redshift subsample into 3 data sets: a training
set, a validation set, and an independent
validation set. The first two sets are required by ANN$z$, and the last
one is used to give an independent photometric redshift validation.

\begin{figure}
\begin{flushleft}
\plotone{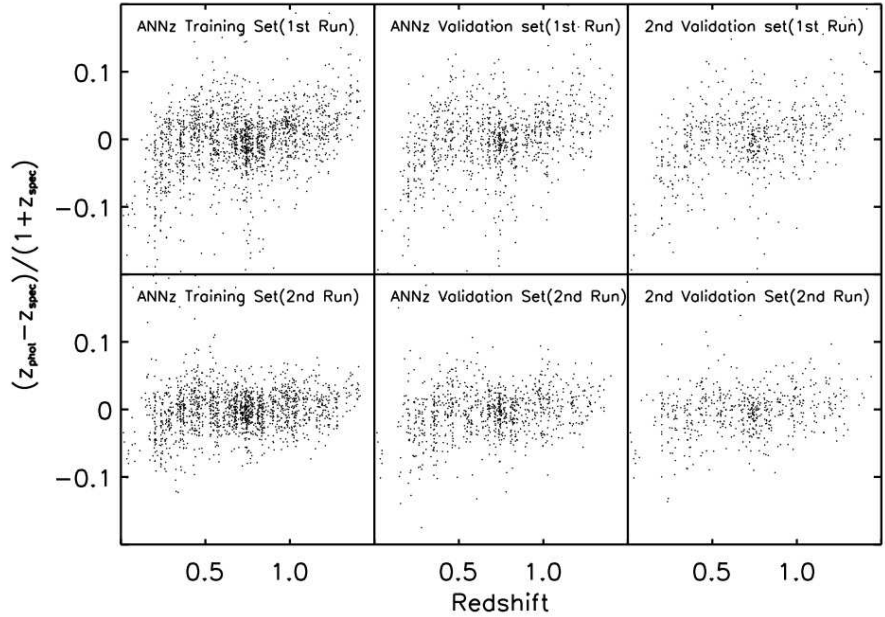}
\end{flushleft}
\caption{A direction comparison between photometric and spectroscopic redshifts.
 Panels left to
  right show the $(z_{\rm phot}-z_{\rm spec})/(1+z_{\rm spec})$ distributions as a function of redshift for the training, validation, and independent
  validation samples respectively.  Upper row shows the first ANN$z$
  estimation run, and the lower row the second run with flagged
  galaxies removed, which shows much less systematic deviations.}
\label{f:zps}
\end{figure}

\begin{figure}
\begin{flushleft}
\plotone{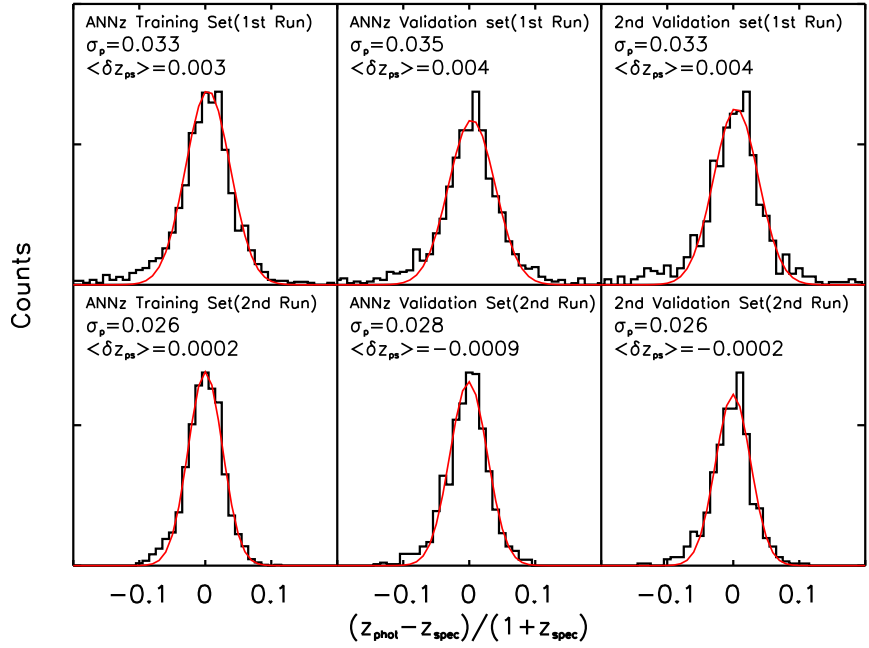}
\end{flushleft}
\caption{Histograms of photometric redshift errors $\delta z_{p}
  \equiv(z_{\rm phot}-z_{\rm spec})/(1+z_{\rm spec})$. Panels left to
  right show errors for the training, validation, and independent
  validation samples respectively.  Upper row shows the first ANN$z$
  estimation run, and the lower row the second run with flagged
  galaxies removed. Gaussian uncertainty $\sigma_p$  and mean $(z_{\rm phot}-z_{\rm spec})/(1+z_{\rm spec})$ for each set are
  indicated in each panel.
\label{f:sigma}}
\end{figure}

Any bad entries, either bad photometry or bad spectroscopic
redshifts, should not be included in the training set.  Including
these outliers will yield large scatters for the resulting
photometric redshifts. We ran the ANN$z$ estimation twice. After the
first run, we compared photometric redshifts with spectroscopic
redshifts in the sample and flagged those with
$|z_p-z_s|/(1+z_s)>3\sigma_p$.\footnote{We define $\sigma_{p}$, the
  calculated uncertainty of the photometric redshifts, as the
  dispersion of $\delta z_{p}\equiv(z_{p}-z_{s})/(1+z_{s})$.}  For a
second training run, we updated the training and validation sets
omitting flagged objects, about 3.5\% of the initial sample.  This
iteration improved the accuracy of the photometric redshifts
significantly as shown in Figure~\ref{f:zps} and Figure~\ref{f:sigma}.


It is critical to assess the accuracy of photometric redshifts before
using them.  Traditionally objects with spectroscopic redshifts are
used to evaluate photometric redshift errors.  ANN$z$ requires one
validation set of objects, and we created an independent subsample
with spectroscopic redshifts, none of which is used for the ANN$z$
training, for an additional validation. The comparison between
photometric and spectroscopic redshifts in Figure~\ref{f:zps} permits a direct estimation
of photometric redshofts errors. The photometric redshifts derived in the first run
show much more systematic deviations caused by the bad redshifts in the training set.
Figure~\ref{f:sigma} shows a
Gaussian distribution of redshift errors, implying a well-defined
random uncertainty characterizes most of the photometric
redshifts. Both the ANN$z$ and independent validation sets yield
similar $\sigma_{p}\sim 0.025$ in the second run, again with about
3.5\% outliers.  The training set $\sigma_p$ cannot be used for an
uncertainty estimate because it was deliberately minimized by the
calculation. The outlier galaxies have a median 3.6\,$\mu$m flux
density of 12\,$\mu$Jy, much higher than the 6.5\,\mmjy median flux
density for the whole sample.  The higher average 3.6\,$\mu$m flux
densities outlier galaxies may have IRAC fluxes contaminated by
nearby objects, or two sources are unresolved by IRAC but resolved at
shorter wavelengths.  The IRAC 3.6\,$\mu$m image has a 2\arcsec\
angular resolution, whereas the CFHT $R$-band images from which the
DEEP2+3 sample was selected have resolution of $\sim$1\arcsec.
Contamination or source blending are therefore expected to occur.

Galaxies with spectroscopic redshifts tend to be the brightest
galaxies in any sample, and this is especially true 
in our 3.6\,$\mu$m-selected sample.  Galaxies that are faint in
visible light  have relatively larger photometric errors and
therefore may have larger errors in their photometric redshifts. Validation
with the spectroscopic redshift subsample may not characterize the
photometric redshift scatter for the fainter galaxies, and it is essential
to assess photometric redshift accuracy for these  faint objects.

We propose a new method to estimate the photometric redshift standard
deviation for the whole sample. The method is based on galaxy-galaxy
pair statistics.  A substantial fraction of galaxy visual pairs are
actually physically related, and the redshift offset for a real pair
should be near zero.  However, the calculated photometric redshift
offset for a real pair may be nonzero if either redshift is erroneous.
We define a visual pair as any two galaxies within an angular
separation of 30\arcsec\ and define the redshift offset as $\delta
z_{\rm ss}\equiv (z_{\rm s1}-z_{\rm s2})/(1+z_{\rm s1})$, where
$z_{\rm s1}$ and $z_{\rm s2}$ are spectroscopic redshifts for two galaxies of the
pair.  The 30\arcsec\ angular distance is corresponding to a distance of 160-240 kpc  for galaxies
at z=0.4-1.2, similar to the criterion of d$<$300kpc adopted
by \citet{lin2007} to study galaxy pairs and mergers in the same redshift range in the same field.
\citet{lin2007}
found that these galaxy pairs have a very wide range of star formation activity.
Galaxy pairs with shorter separation ($\sim$50kpc) likely have higher star formation activity,
otherwise they do not show any bias towards any type. 

\begin{figure}
\plotone{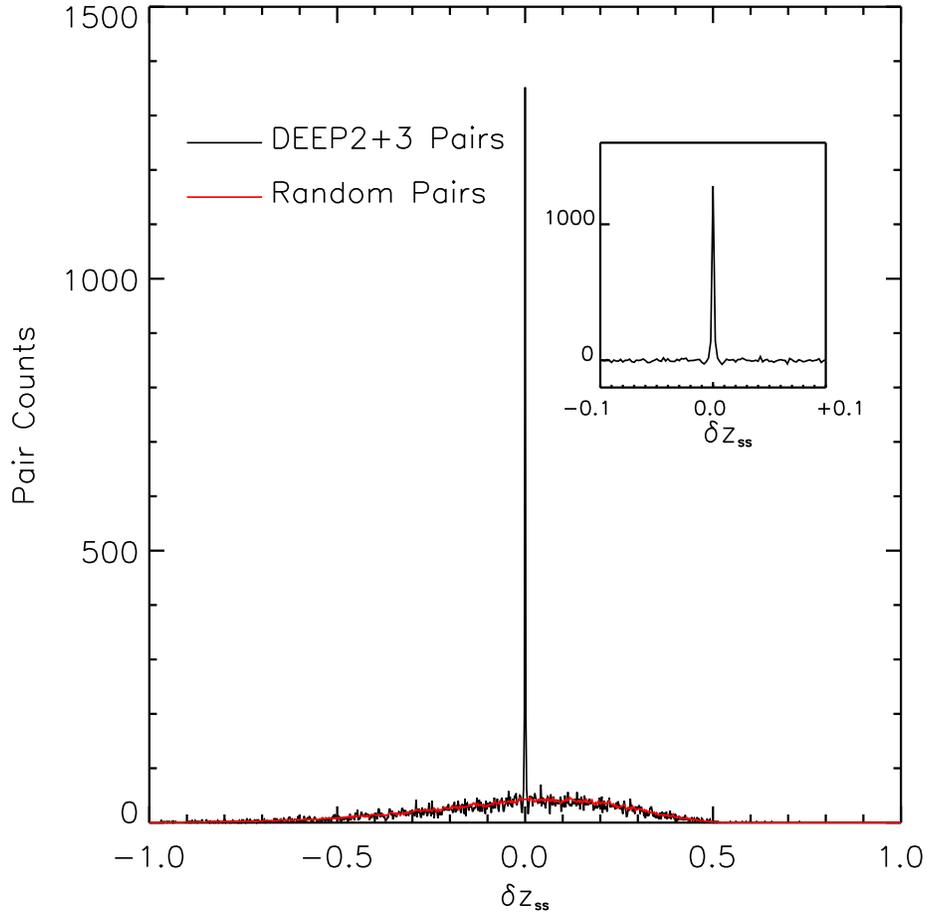}
\caption{Histogram of normalized redshift offsets for galaxy pairs
  selected from the spectroscopic subsample.  The red line shows the
  distribution expected for random, unrelated galaxies as calculated
  by a Monte-Carlo simulation.  The insert shows the residual
  real pair distribution after the random distribution of unrelated
  pairs is subtracted.}
\label{f:dz_ss}
\end{figure}

Figure~\ref{f:dz_ss} show the histogram of $\delta z_{\rm ss}$ 
for all visual pairs with both galaxies having redshifts in
the DEEP2+3 sample.  The histogram shows two components: a spike at
$\delta z_{\rm ss}=0$ for real pairs and a broad distribution for
projected pairs that are physically unrelated. The expected $\delta
z_{\rm ss}$ for projected pairs can be easily generated using
Monte-Carlo simulation and subtracted from the distribution for all
pairs. The residual spike is the $\delta z_{\rm ss}$ distribution for
real pairs.  As
expected, both of these velocity components are small compared to the
expected uncertainty in the photometric redshifts.

\begin{figure}
\plotone{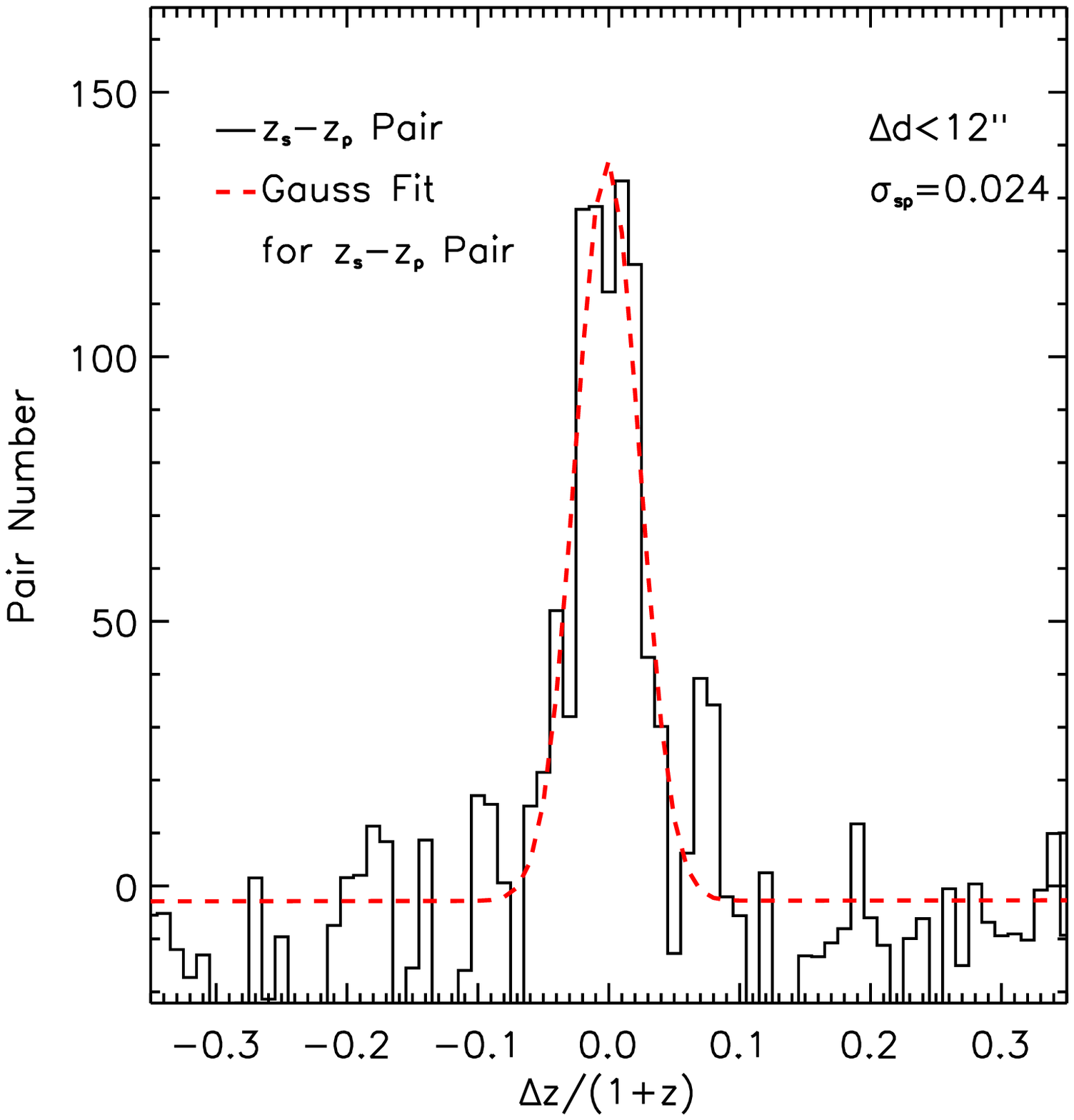}
\caption{Histogram of normalized redshift offsets for galaxy pairs
  with one photometric and one spectroscopic redshift. The redshift offset distribution for the random, unrelated galaxy pair is already subtracted
with the Monte-Carlo simulation in this diagram. 
The red dashed line shows a Gaussian fit to the
  real pair distribution.}
\label{f:ddz_sp}
\end{figure}

Having a large number of spectroscopic redshifts available allows
pair statistics to give a good estimate of the uncertainty in the
ANN$z$ redshifts.  The  pair sample tested consists of galaxies separated by
$<$30\arcsec\ where one galaxy has a spectroscopic redshift and the
other only a photometric redshift.  The normalized redshift offset is
defined as $\delta z_{\rm sp}\equiv (z_p-z_s)/(1+z_s)$, where $z_p$
is the photometric redshift and $z_s$ is the spectroscopic redshift
of the respective galaxies. The histogram for all pairs
(Figure~\ref{f:ddz_sp}) shows two components, as with spectroscopic
pairs. Subtracting the simulated distribution for unrelated pairs
from the real pair distribution gives the $\delta z_{sp}$
distribution for physical pairs.  This distribution shows a Gaussian
shape (Figure~\ref{f:ddz_sp}) with $\sigma_{\rm sp}=0.024$. This is
very close to $\sigma_p$, the standard deviation of the photometric
redshifts estimated from the difference between spectroscopic and
photometric redshifts for the same galaxies.

\begin{figure}
\plotone{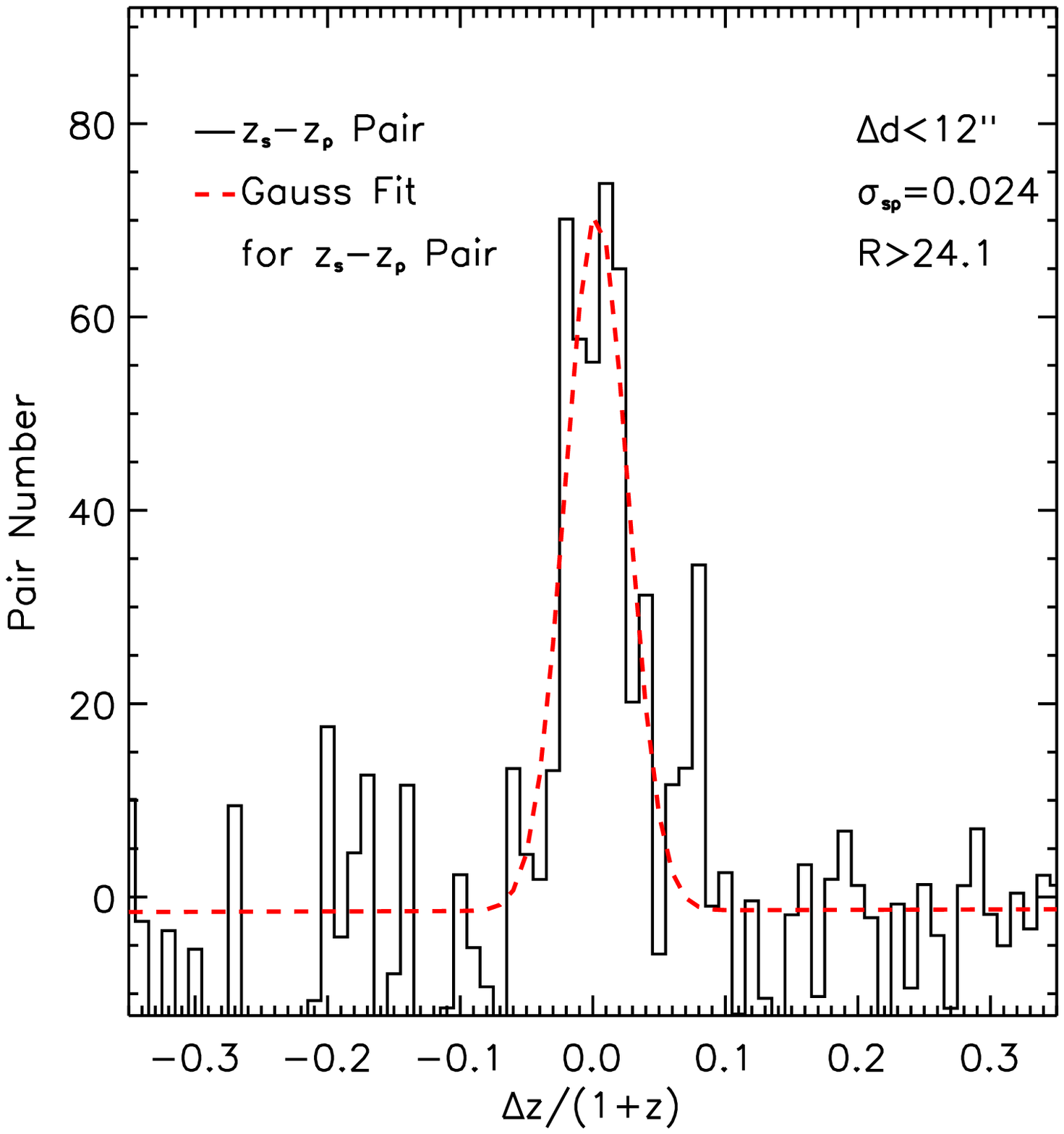}
\caption{Histogram of normalized redshift offsets for galaxy pairs
  with one photometric and one spectroscopic redshift with the
  photometric component having $R>24.1$. The redshift offset distribution for the random, unrelated galaxy pair is already subtracted
with the Monte-Carlo simulation in this diagram. The red dashed
  line shows a Gaussian fit to the real pair distribution.}
\label{f:dz_sp_faint}
\end{figure}

The pair method can be used directly to validate the photometric
redshifts for faint galaxies with $R>24.1$. Figure~\ref{f:dz_sp_faint}
shows the results.  $\delta z_{sp}$ for these galaxies also shows a
Gaussian shape with $\sigma_{\rm sp}=0.024$. The pair statistics thus confirm the
validity of the ANN$z$ photometric redshifts even for galaxies
fainter than the limiting magnitude of the training set.

Overall, $\sigma_p$ of the photometric redshifts for the whole 3.6\,\micron\
sample is very close to that of recent state-of-art SED-fitting
photometric redshifts for optically-selected samples, which often
have  more than 20 bands of
intermediate and broadband photometry
\citep{wol03,ilbert2006,ilbert2009, whitaker2011}. Comparing with
those photometric redshift estimations, the ANN$z$ redshifts
have well-defined and consistent error distributions for both
optically bright and faint galaxies but are inherently limited to
$z\le1.25$, the maximum redshift of the training set. 

\section{Galaxy Populations in the 3.6\,$\mu$m Selected Sample}

\subsection{Color Classification}

\begin{figure}
\epsscale{0.8}
\plotone{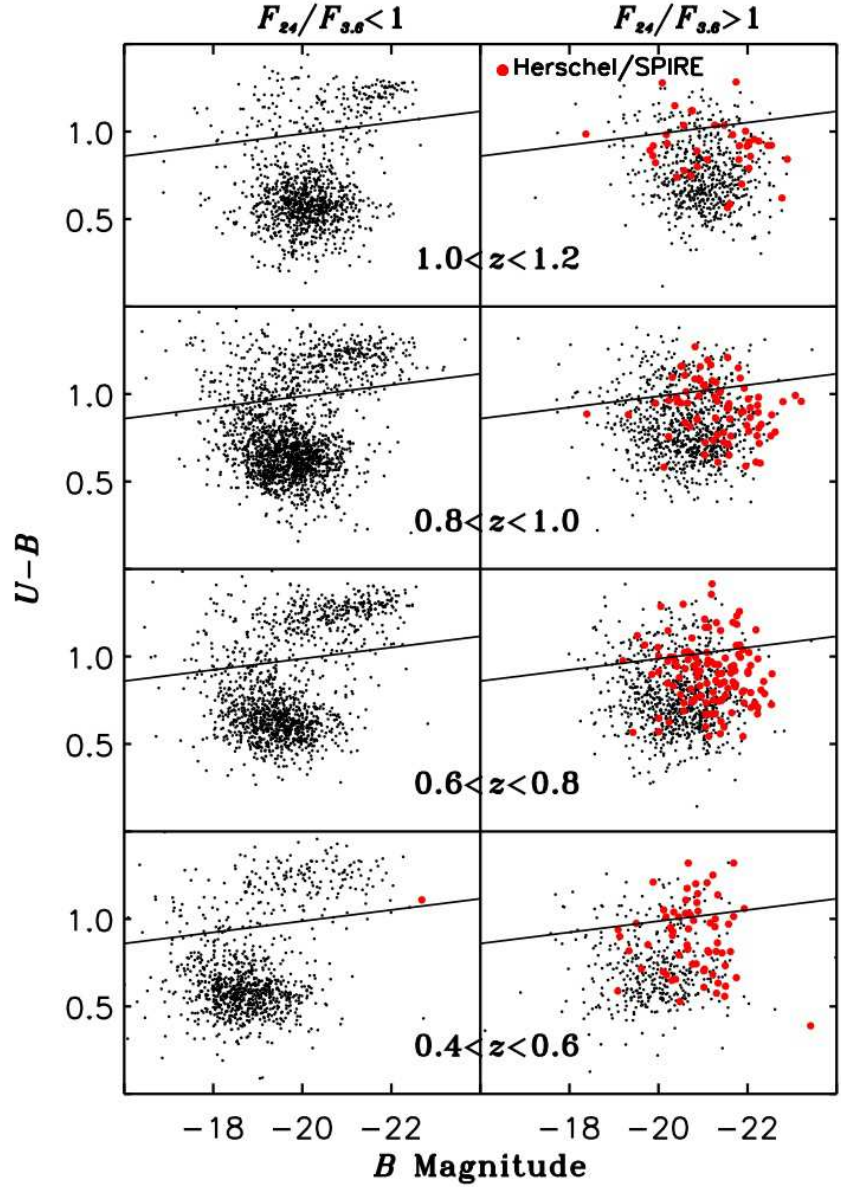}
\caption{$U-B$ vs $M_{B}$ color-magnitude diagram for the
  3.6\,$\mu$m selected sample in  redshift
  bins as indicated in the figure. The left panels are for non-dusty galaxies
  ($f_{24}/f_{3.6}<1$), and right panels are for dusty galaxies
  ($f_{24}/f_{3.6}>1$). The red dots denote \h/SPIRE sources
  selected at 250\,$\mu$m \citep{rigopoulou2012}. Most of these
  sources are LIRGs or ULIRGs, and even some of the LIRGs and ULIRGs
  are in the red sequence.}
\label{f:ub_b}
\end{figure}

With redshifts known, deriving rest frame colors and magnitudes for
sample galaxies is straightforward.  All types of galaxies have very
similar SEDs at NIR wavelengths.  The $K$-band absolute magnitudes
for galaxies can therefore be calculated with the 3.6\,\micron\ flux
densities and K-correction from \citet{depropris2007}. The absolute
magnitude range for the sample is $-19<M_K<-25$.  Rest frame $U-B$
colors for the sample were derived from $V-I$ colors via a set of
local galaxy SED templates (\citealt{kinney1996} ---
\citealt{willmer2006} and \citealt{faber2007} took a similar
approach).  Rest frame absolute $B$ magnitudes $M_B$ were derived in
the same way.  The RS and BC
\citep{pvd00} were divided with the \citet{willmer2006} criterion:
\begin{equation} (U-B)_0=-0.032(M_B+21.52)+0.454-0.25+0.832
\label{eq:sep}
\end{equation} 
using the same $U$ and $B$ magnitude system they used
but adding 0.832 to convert  from the Vega to the AB system. 
Figure~\ref{f:ub_b} shows a series of color-magnitude (CM) diagrams for the
3.6\,$\mu$m-selected sample.

\begin{figure}
\plotone{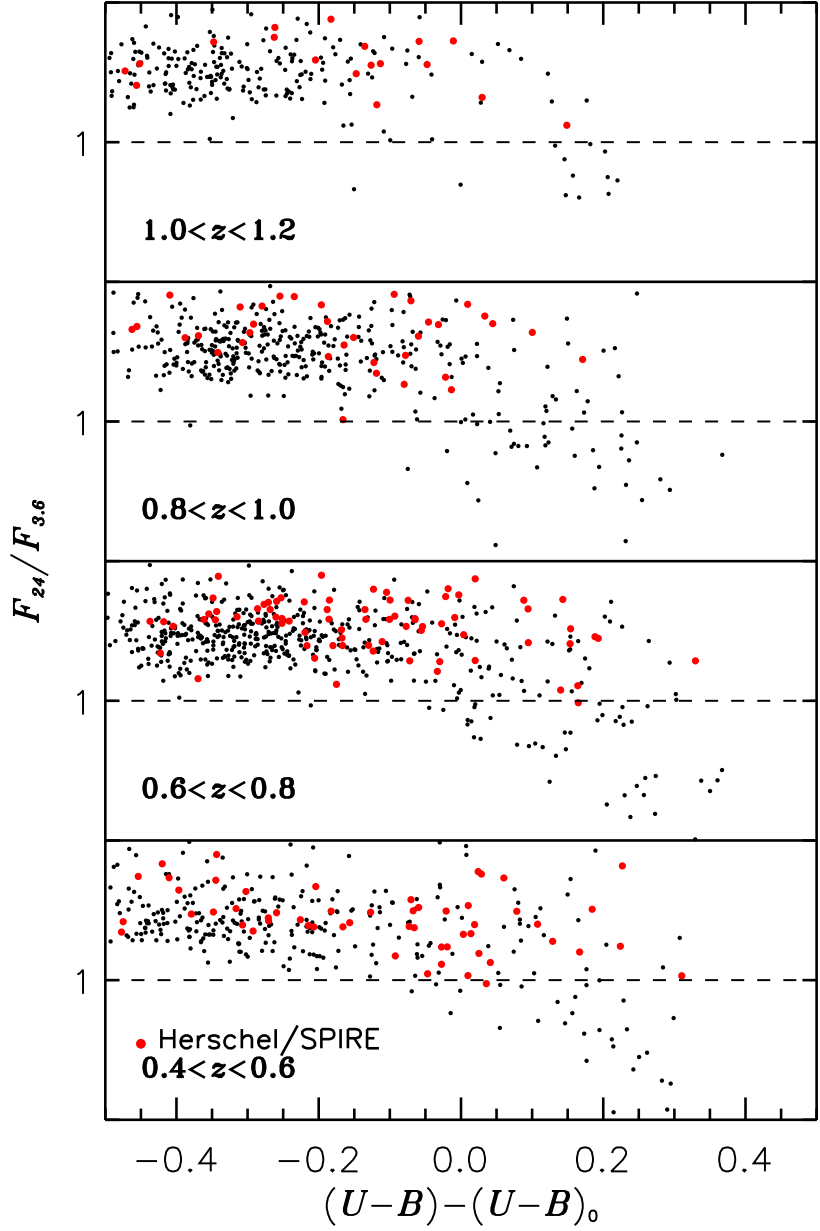}
\caption{Infrared-visible color-color diagram for all galaxies in the
  3.6\,\micron\ sample detected at 24\,$\mu$m.  Redshift bins are
  indicated.  The abscissa is the $U-B$ color of each galaxy minus
  $(U-B)_0$ as given by equation~\ref{eq:sep}. Red dots denote
  \h/SPIRE sources detected at 250\,$\mu$m \citep{rigopoulou2012}.}
\label{f:f24f36ub}
\end{figure}

Two types of galaxies have colors placing them in the RS:
dusty star-forming galaxies and quiescent galaxies with little or no
star formation\citep{labbe2005}.  Current studies use either a rest-frame
visible color-color diagram or IR emission to separate these types
\citep[e.g.,][]{brammer2009, williams2009,
  wang2012}. \citet{williams2009} showed that the types can be
separated in a $U-V$ vs $V-J$ diagram, where dusty galaxies have
redder $V-J$ colors than quiescent galaxies. \citet{brammer2009} and
\citet{wang2012} used 24\,$\mu$m emission to identify dusty star
forming galaxies.  The 24-to-3.6\,\micron\ flux density ratio,
$f_{24}/f_{3.6}$, is closely related to the specific star formation
rate and should therefore provide a good way of distinguishing
passive from dusty star-forming galaxies.  Figure~\ref{f:f24f36ub}
shows that $f_{24}/f_{3.6}$ is roughly constant for 24\,$\mu$m
sources with BC $U-B$ colors, but this ratio decreases to
below 1 for 24\,$\mu$m sources with RS colors. Nearly all
24\,$\mu$m galaxies in the BC have $f_{24}/f_{3.6} >1$,
whereas more than 50\% of 24\,$\mu$m galaxies in the RS
have $f_{24}/f_{3.6} <1$.  All \h/SPIRE sources in the RS
also have $f_{24}/f_{3.6} >1$.  These sources must be star forming to
be detected with SPIRE.  We have therefore used $f_{24}/f_{3.6} =1$
to separate dusty from passive galaxies in the RS: galaxies
with $f_{24}/f_{3.6} <1$ are taken to be passive galaxies with
24\,$\mu$m residual emission, and galaxies with $f_{24}/f_{3.6} >1$
in the RS are taken to be dusty and star forming.  This
criterion works because most RS galaxies in the sample are very
luminous with $f_{3.6} >20$\,\mmjy\, and therefore RS galaxies not detected at
24\,$\mu$m must have $f_{24}/f_{3.6} <1$.



Galaxies with colors near the dividing line between the RS and the BC
are called ``green valley'' objects. Most green valley galaxies are
also MIPS 24\,$\mu$m sources \citep{cowie2008,kartaltepe2010}. An
IR-selected sample has many more green valley objects than a
visible-selected sample, and this is the case for our
3.6\,$\mu$m-selected sample. Figure~\ref{f:ub_b} shows many MIPS
24\,$\mu$m sources in the green valley, and some are well into the RS
color zone.  On the other hand, most galaxies with no MIPS 24\micron\
detection have rest $U-B$ colors well away from the green valley, and
the rest $U-B$ colors should accurately classify these galaxies as
passive or star-forming.


\subsection{Star Formation and Stellar Mass Assembly in the
3.6\,\micron\ Sample}

Stellar mass for the 3.6\,\micron\ sample can be derived from 
stellar population models. In practice we used BC03 \citep{bc03}
models with solar metallicity and Salpeter initial mass function (IMF). The
model template set was generated with a grid of stellar masses,
$E(B-V)$, ages, and e-folding times for star formation history. Stellar
masses for the sample galaxies were determined by fitting their
observed SEDs to the model templates. 
The derived stellar masses 
can be converted to those with the Chabrier IMF \citep{bundy2006} by
adding $-0.25$\,dex and to those with the ``diet Salpeter'' IMF
\citep{bell2003} by adding $-0.15$\,dex. 


Even without detailed modeling, the rest-frame $K$-band luminosity
for a galaxy is a good tracer of its stellar mass
\citep[e.g.,][]{bell2003}. The IRAC 3.6\,$\mu$m band measures
rest-frame $K$-band for a galaxy at $0.4<z<1.2$. Figure~\ref{f:ml}
shows the $K$-band mass/light ratio for the galaxies in four redshift
bins. The mass/light ratio is not constant but depends on color (as
also found by \citealt{bell2003}) and also slightly depends on
redshift. Because the galaxies are bimodal in color, $M/L$ also shows
a bimodality: $M/L\sim 1.25$ for BC galaxies, and $M/L\sim 3$ for RS
galaxies. The different mass-light ratios imply different stellar
mass limits in the 3.6\,$\mu$m-selected
sample for the two galaxy classes.

Figure~\ref{f:cm_mass} shows galaxy color versus mass.  In general,
the most massive galaxies at any redshift are the red, quiescent
ones.  Those are followed by dusty star-forming galaxies, which are
more massive than the remaining star-forming galaxies in the BC. Only
the dusty star-forming galaxies are massive enough to be direct
progenitors of massive red quiescent galaxies. Figure~\ref{f:cm_mass}
also shows that there is deficit of lower-mass RS galaxies at
$1.0<z<1.2$.  This is not a selection effect: galaxies with
$M\approx3\times10^9$\,\Msun\ would be readily detectable regardless
of color.  The deficit of low mass galaxies on the RS at high
redshift shows that the most massive RS galaxies formed first, many
of them at $z>1.2$. The high redshift \h/SPIRE sources, representing
LIRGs and ULIRGs, tend to be massive, indicating that intensive star
formation occurs in massive galaxies at these redshifts. 



Figure~\ref{f:ub_z} shows galaxy color versus redshift in six stellar
mass bins. Most galaxies in the highest mass bin are quiescent,
indicating that they were already in place at $z=1.2$. There are a few star
forming galaxies  in this mass bin as well. On the other hand, the
most intensive star formation was occurring in massive galaxies with
$M_*>10^{10}$\,M$_{\odot}$ at $0.4<z<1.2$. Figure~\ref{f:sfr_mass}
shows total infrared luminosities against stellar mass. The plot
shows that most of the SPIRE sources are LIRGs or ULIRGs, and most
are in the green valley \citep{rigopoulou2012}. Figure~\ref{f:sfr_mass} is very similar to
Fig.~1 of \citet{noekse2007} for the visible-selected spectroscopic
sample in the same field, showing that more massive
galaxies have higher total infrared luminosities. Yet the Herschel SPIRE selected galaxies are biased towards
the high L$_{\rm IR}$, comparing with the mean SFR-stellar-mass relation in the same redshift range for a more sensitive 24$\mu$m selected sample\citep{elbaz2007}. All of the ULIRGs are more massive than 10$^{10.5}$\,\Msun. 
The mean stellar mass for
LIRGs/ULIRGs in the sample decreases from $10^{10.72}$\,M$_{\odot}$
at $z=1.1$ to $10^{10.50}$\,M$_{\odot}$ at $z=0.5$, suggesting that
more massive galaxies migrated to the RS earlier through
intensive star formation.


Figure~\ref{f:ub_z} also shows that quiescent galaxies at $z>1.1$ are absent in the
$10<\log(M_*/\Msun)<10.5$ bin and so are those at $z>1$ in the the
$9.5<\log(M_*/\Msun)<10$ bin. This suggests that quiescent galaxies
with low stellar mass did not form until low redshifts. Figure~\ref{f:smass_czr} shows the RS mass fraction as a
function of redshift in each stellar bin. Though the mass fraction of
quiescent galaxies in $11<\log(M_*/\Msun)<11.5$ bin has large error
bars, it is $\sim$75\% and roughly constant from $z=0$ to $z=1$. This
suggests that massive galaxies had largely stopped forming stars by
$z=1.2$. At $10.5<\log(M_*/\Msun)<11$, the quiescent fraction
increases from 40\% at $z=1.1$ to 60\% at $z=0.6$, close to the local
value \citep[58\%,][]{bell2003} at $z\sim0$. In the
$10<\log(M_*/\Msun)<10.5$ bin, the mass fraction increases from 10\%
at $z=1.1$ to 30\% at $z=0.5$, also close to the local value of
41\%. The x-intercept of this distribution is at $z\sim1.3$. Quiescent galaxies in
$9.5<\log(M_*/\Msun)<10$ are a small fraction of the total at all
epochs.  The first such galaxies appeared around $z\approx1.1$, and
their numbers have increased to about 25\% at $z\approx0$. 

There are two possible causes for the deficit of red quiescent galaxies in lower mass bins in
Figure~\ref{f:smass_czr}: (1) galaxies with lower mass migrated to the RS later;
or (2) quienscent galaxies were undergoing either dry or mixed merging with high rate to move them quickly
from lower mass end to high mass end in the RS\citep{faber2007}.
The merging scenario requires such a high merging rate in low mass bins that galaxies just migrating to the RS were immediately undergoing dry/mixed merging. Yet \citet{lin2008} found
 that fractions of dry and mixed mergers at z$\sim$1.1 are only 8\% and 24\%  for galaxies with a typical stellar mass
of 2$\times$10$^10$ M$_{\odot}$. It is unlikely that dry and mixed merging rate for galaxies with lower stellar masses
would increase and be the major cause for the deficit of red quiescent galaxies in lower mass at z$\sim$1.

In general, massive quiescent galaxies were already in place by $z=1.2$, and  quiescent galaxies with $9.5<\log(M_*/\Msun)<10.5$ started to form
in 1.1$<$z$<$1.3. This indicates that the quenching of star formation, regardless of types, occurs earlier in massive galaxies
later in galaxies with lower stellar masses.

\subsection{$K$-band Luminosity Function and Stellar Mass Functions}

The rest $K$-band luminosity function is an important measurement of
galaxy population evolution. It is equivalent to a stellar mass
function but with a better defined completeness limit and much less
model dependence. This is particularly true for our 3.6\,$\mu$m
selected sample. Because the IRAC 3.6\,\micron\ probes roughly the
rest-frame $K$-band for galaxies at $0.4<z<1.2$, the
3.6\,$\mu$m-to-$K$ K-correction is effectively independent of galaxy color,
permitting an unbiased comparison of the luminosity functions at
different redshifts. We have derived the $K$-band luminosity
functions for quiescent, star-forming, and all galaxies in the EGS
field at $z=0.5$, 0.7, 0.9, and 1.1 with two methods: parametric and
non-parametric. The non-parametric method is the familiar $1/V_{max}$
\citep{fel76, eales1993}, and the parametric method is ``STY'' \citep{sty75},
which fits a Schechter function to the sample. The STY method itself
determines only $\alpha$ and $M_K^*$ in the Schechter function. The
normalization factor $\phi^*$ was derived using the minimum variance
density estimator \citep{HD1982}.
Both  luminosity functions  are plotted in
Figures~\ref{f:klf}, and numerical values are given in
Table~2.


The major uncertainty in a luminosity function estimate based on a
sample with small coverage or a single field is cosmic variance due
to galaxy clustering \citep{hua97,som04, newman2002}.  Our $K$-band
luminosity functions were derived from a single
1\arcdeg$\times$10\arcmin~field. The maximum angular scale for this
sample is 1\arcdeg, and thus the cosmic variance values, $\sigma_v$,
at $z=0.5$--1.1 are 28\%--36\% assuming a general correlation length
$r_0=5$\,Mpc \citep{som04}. NIR-selected galaxies have a correlation
length similar to that of visible-selected samples:
\citet{roche2003} and \citet{oliver2004} measured correlation lengths
of 4.1 and 6.4\,Mpc for $K$- and IRAC-selected samples,
respectively. Recently \citet{grazian2006} and \citet{quadri2007}
measured two-point correlation functions for distant red galaxies
(DRGs) at $z\sim2$ and obtained $r_0=7$--11\,Mpc. The DRGs have a
typical stellar of $10^{11}$\,M$_{\odot}$. If we adopted the largest
$r_0=13.5$, the cosmic variance would be 2.5 times larger.
However, this
large correlation length is not applicable to the bulk of our sample
but only to the most massive galaxies. The cosmic variance can also
be calculated using dark matter variance and bias factor
\citep{roche2003}. The volumes for our sample in 4 redshift bins are
$\sim$1.2--$2.8\times10^5$\,Mpc$^3$, and within these volumes the
dark matter fluctuation $\sigma_{DM}\approx0.25$.  The number densities
for $L^*$ galaxies in our sample are $10^3$\,Mpc$^{-3}$
(Figure~\ref{f:klf}), and therefore the bias for this population at
$z\sim1$ is $b\sim1.5$. According to \citet{roche2003}, the cosmic
variance $\sigma_v=b\sigma_{\rm DM}$, and $\sigma_v \sim0.3$--0.4 for
the $L_*$ galaxies in our sample, consistent with values above. Thus all
methods suggest that the variance for $L^*$ galaxies in this sample
is about 30\%.

$B$-band luminosity functions for galaxies at $0.4<z<1.2$ are well
established both in the EGS and a wide range of other fields
\citep{willmer2006,faber2007}. The comparison can be used to correct
the EGS LF to the global average.  The 3.6\,$\mu$m sample is
incomplete at faint $B$ magnitudes (Figure~\ref{f:blf1}). The
3.6\,$\mu$m limiting magnitude is 23.2, corresponding to absolute
magnitudes  $M_K\approx-20$ to $-21$ at $0.4<z<1.2$.  The rest-frame
$B-K$ colors for galaxies are in the range of $-1<B-K<2$, giving a
$B$-band completeness limit of $-20.5$ to $-21.0$. The mean density
ratio between the EGS and global $B$-band LFs, taking into account
only galaxies brighter than the completeness limit in each redshift bin, is
$\langle\phi_{\rm egs}/\phi\rangle=0.83$, 1.7, 1.66, and 1.20 for bins at
$z=0.5$, 0.7, 0.9, and 1.1 respectively. The standard deviation of
these values is 0.38, consistent with the theoretical cosmic-variance
estimation for this sample.  The four values of $\langle\phi_{\rm
  egs}/\phi\rangle$ have been applied to the $K$-band LF
and stellar mass function calculated for the 3.6\,\micron\
sample.   The final LFs are plotted in Figure~\ref{f:klf}, and their
values are given in Table~2. These corrected values should be
representative of the entire Universe, not the EGS field itself.

The VIMOS VLT Deep Survey (VVDS) \citep{lefevre2005}  provides
independent  $K$-band luminosity functions at $z=0.5$, 0.7, 0.9, and
1.1.  The VVDS itself measured redshifts in a wide area 
also covered by the \s\ Wide-Area Infrared Extragalactic Survey
(SWIRE) \citep{Lonsdale2004}. \citet{arnouts2007} used
these data to derive $K$-band luminosity functions in the same
redshift bins we have used. Our $K$-band LFs are generally consistent
with those of \citeauthor{arnouts2007}, but the VVDS LFs at $z=0.9$
and 1.1 are 2 times higher than ours at luminosities below $L_*$. We
show later that the VVDS sample has higher mass densities for
star-forming galaxies at $0.4<z<1.2$ than those obtained in any other
survey and argue that the excess star-forming galaxies are the cause
of their high $K$-band luminosity functions at the faint end. Our
$K$-band luminosity functions and the VVDS ones with $L>L^*$ are
higher then the local $K$-band luminosity function \citep{huang2003,
  jones2006}. This is not due to number evolution for massive
galaxies because massive galaxies can only increase in number towards
lower redshifts.  Thus it implies an luminosity evolution in the
$K$-band from $z\sim1$ to the present epoch. 

\begin{figure}
\epsscale{1.0}
\plotone{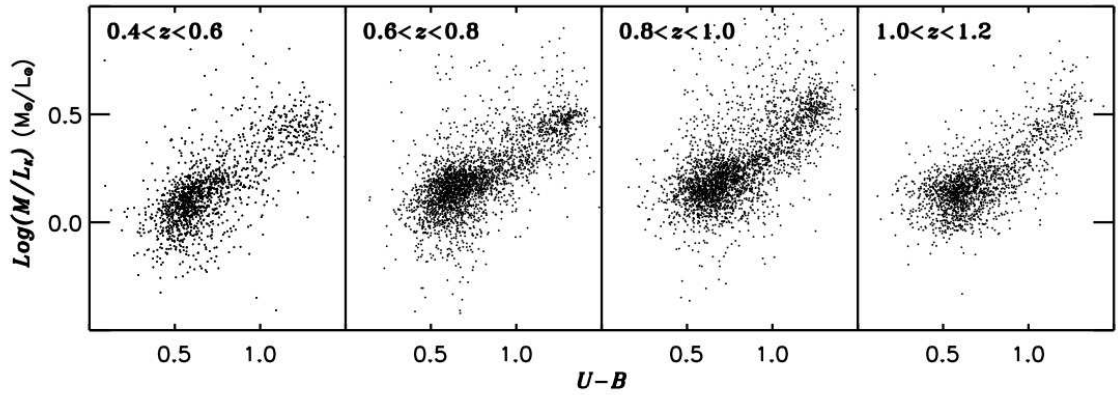}
\caption{$K$-band mass/light ratio versus $U-B$ color.  Points denote
  all galaxies in the 3.6\,$\mu$m-selected sample. Redshift bins are
  indicated in each panel.}
\label{f:ml}
\end{figure}
\begin{figure}
\epsscale{0.8}
\plotone{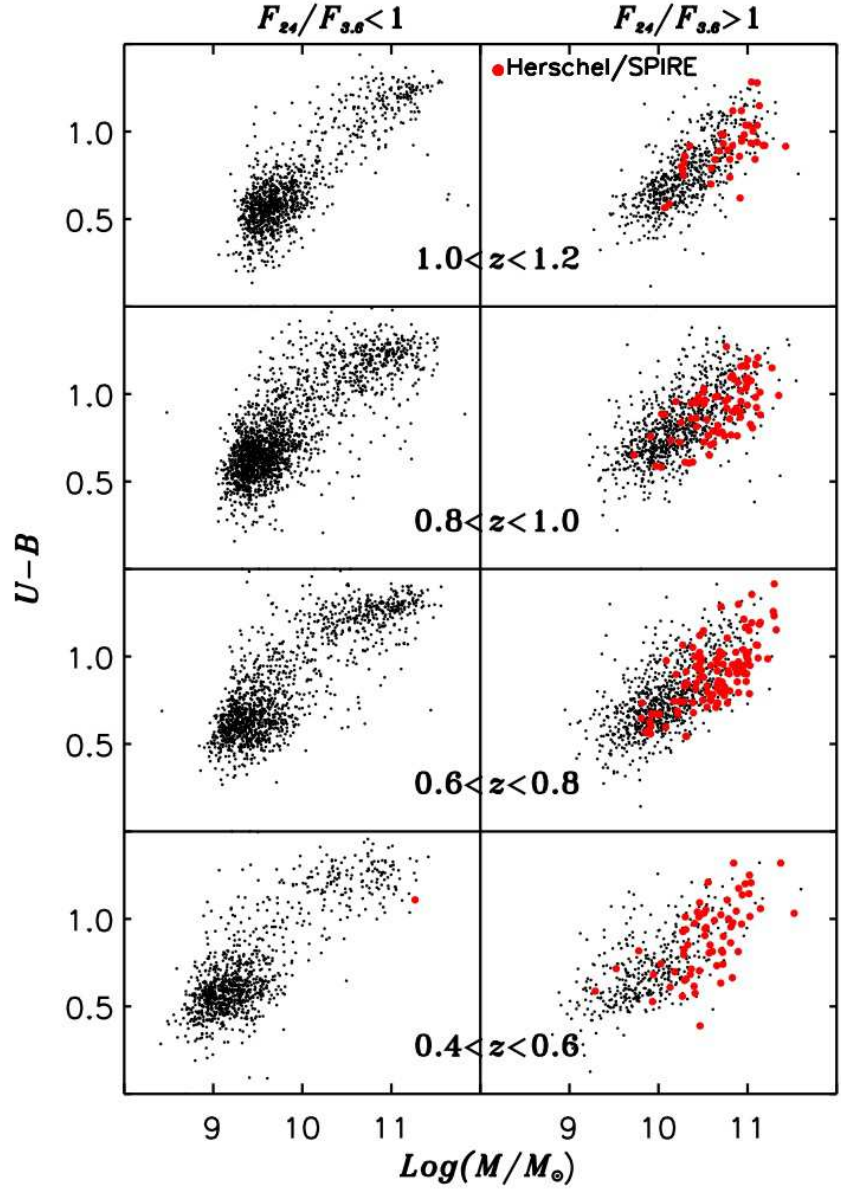}
\caption{Color versus stellar mass for the 3.6\,$\mu$m sample. Redshift bins
  are indicated. The left panels are for non-dusty galaxies
  ($f_{24}/f_{3.6}<1$), and right panels are for dusty galaxies
  ($f_{24}/f_{3.6}>1$). The red dots denote \h/SPIRE sources selected
  at 250\,$\mu$m \citep{rigopoulou2012}.}
\label{f:cm_mass}
\end{figure}
\begin{figure}
\epsscale{1.0}
\plotone{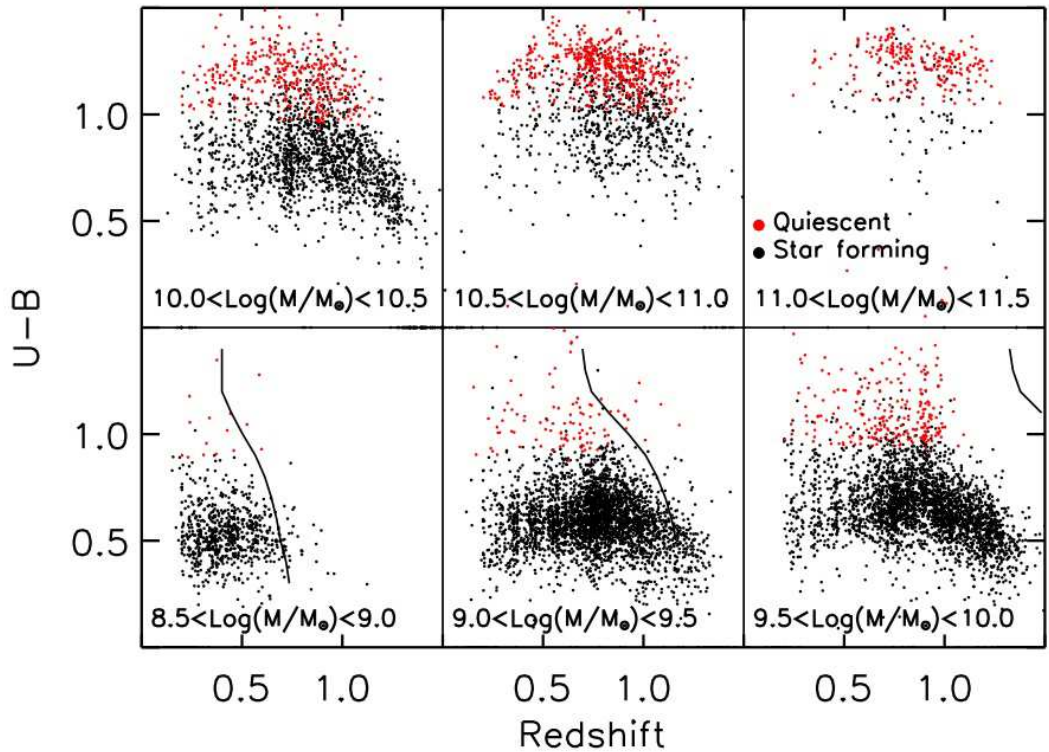}
\caption{Color versus redshift for the 3.6\,$\mu$m sample.  Stellar
  mass bins are indicated in each panel. The thick lines in the three
  lower panels indicate redshift limits to detect galaxies in these mass bins with typical
  colors. The limit ($z>1.5$) is outside the diagram in the upper
  panels. Quiescent galaxies are absent at high redshift end in 3 middle mass bins.}
\label{f:ub_z}
\end{figure}
\begin{figure}
\plotone{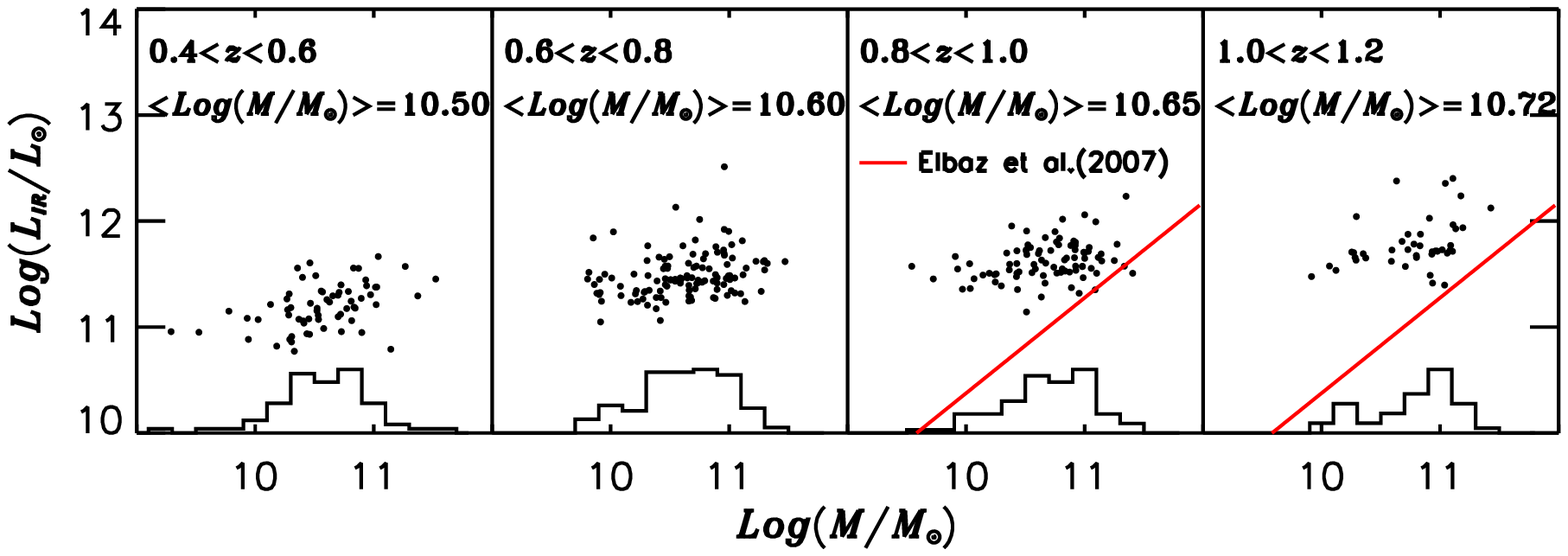}
\caption{$L_{\rm IR}$-stellar-mass relation for the \h/SPIRE sources
  in the 3.6\,\micron\ sample.  Points denote individual galaxies
  with $L_{\rm IR}$ values from \citet{rigopoulou2012}.  The red lines are the mean SFR-stellar-mass relation
(converted to $L_{\rm IR}$) for GOODS galaxies at 0.8$<$z$<$1.2\citep{elbaz2007}. The Herschel galaxies are
strongly biased towards the higher luminosity side. Redshift
  bins are indicated in each panel. Histograms show the stellar mass
  distribution of the SPIRE sources, and their median stellar mass is
  noted in each redshift bin.}
\label{f:sfr_mass}
\end{figure}
\begin{figure}
\plotone{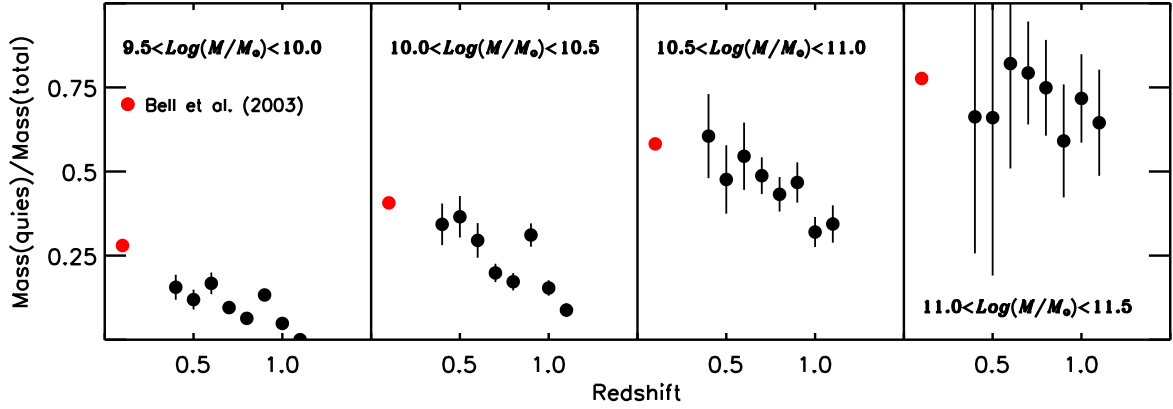}
\caption{Fraction of quiescent galaxies as a function of
  redshift. Mass range is indicated for each panel. Black points
  denote the average mass fraction in small redshift ranges for the
  3.6\,\micron\ sample, and red points denote the local value from
  \citet{bell2003}.}
\label{f:smass_czr}
\end{figure}
\begin{figure}
\epsscale{0.8}
\plotone{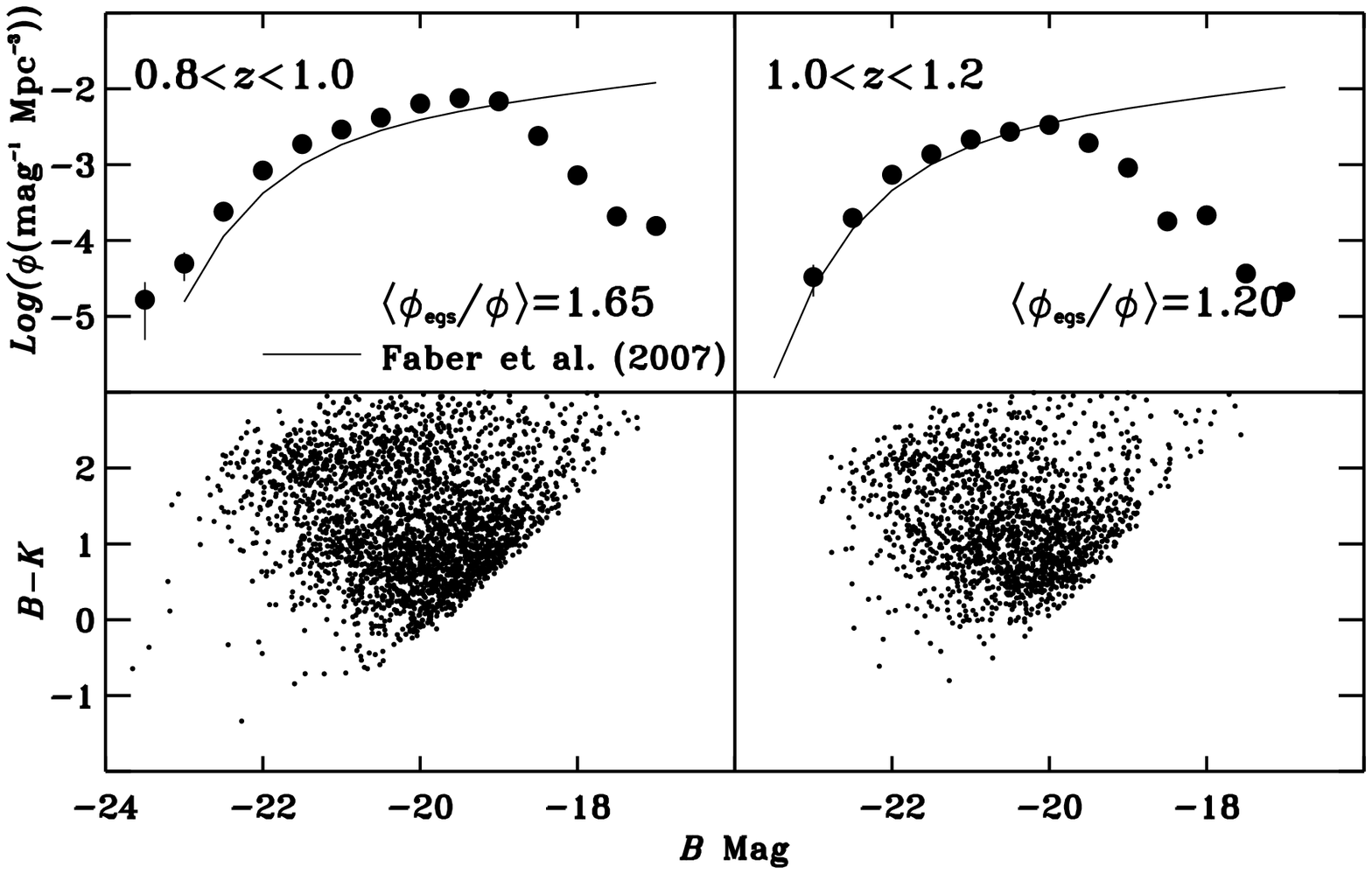}
\plotone{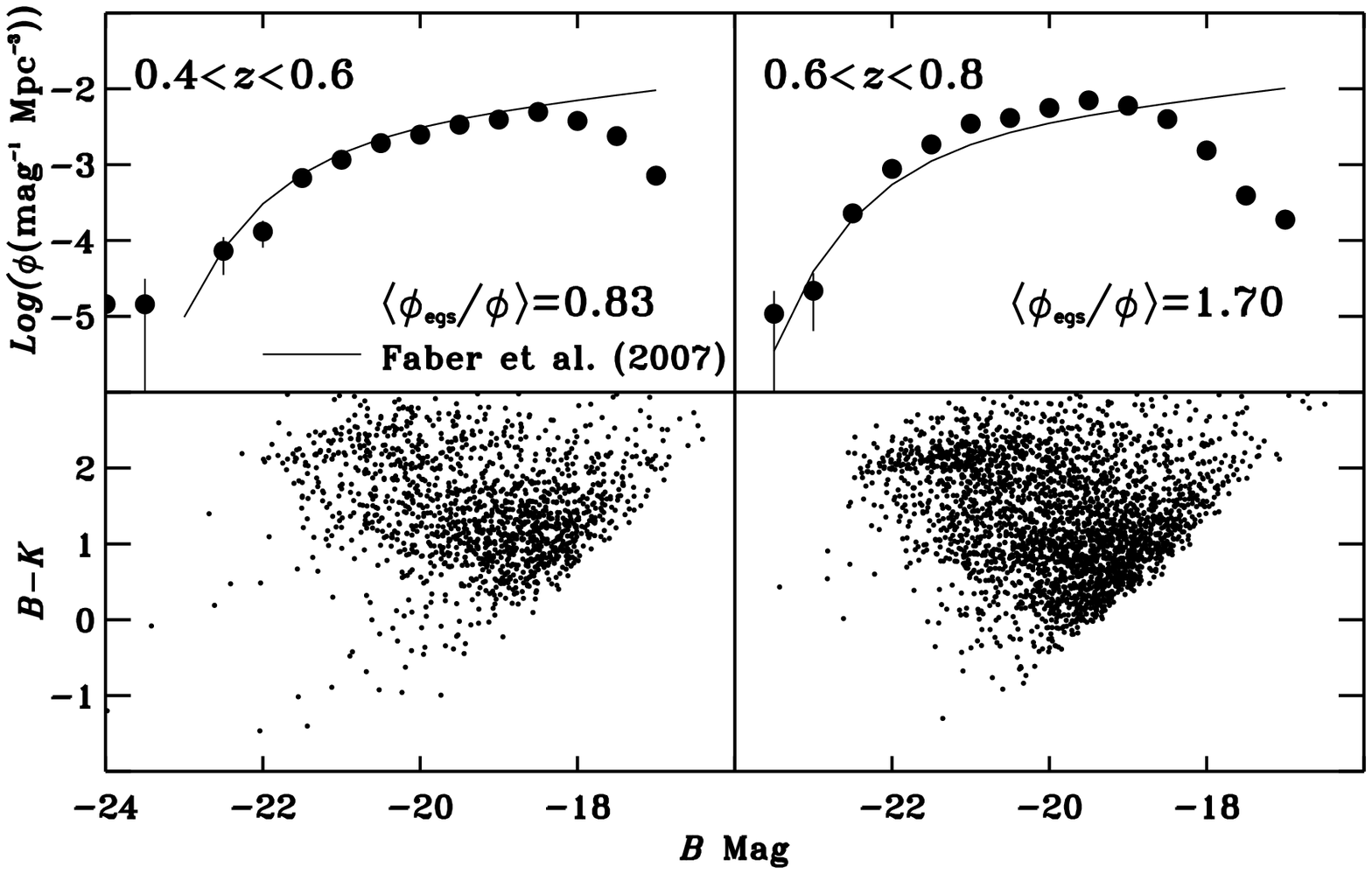}
\caption{$B$-band luminosity functions (upper panels) and
  color-magnitude diagrams (lower panels).  Redshift ranges are
  indicated.  Points in upper panels denote the luminosity function
  derived for the 3.6\,$\mu$m sample, and lines show the luminosity
  function derived by \citet{faber2007} from a visible sample. Points in
  lower panels show color and magnitude for 3.6\,\micron-selected
  galaxies.  The absence of points in the lower right of the lower
  panels shows the incompleteness of the 3.6\,\micron\ sample for
  faint, blue galaxies.}
\label{f:blf1}
\end{figure}
\begin{figure}
\epsscale{1}
\plotone{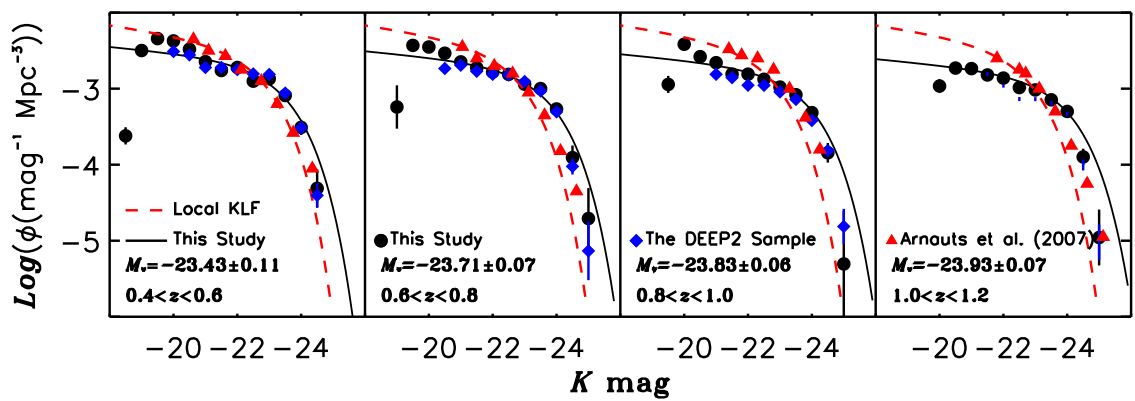}
\caption{$K$-band luminosity functions. Blue points and solid lines
  show results from this study, while red points and dashed lines
  show comparisons from \citet{jones2006} for the leftmost panel and
  \citet{arnouts2007} for the other three panels.   Redshift bins are
  indicated in each panel.  The lines represent  Schechter functions,
  parameters of which  are given in Table-\ref{t:klf}.}
\label{f:klf}
\end{figure}
\begin{figure}
\plotone{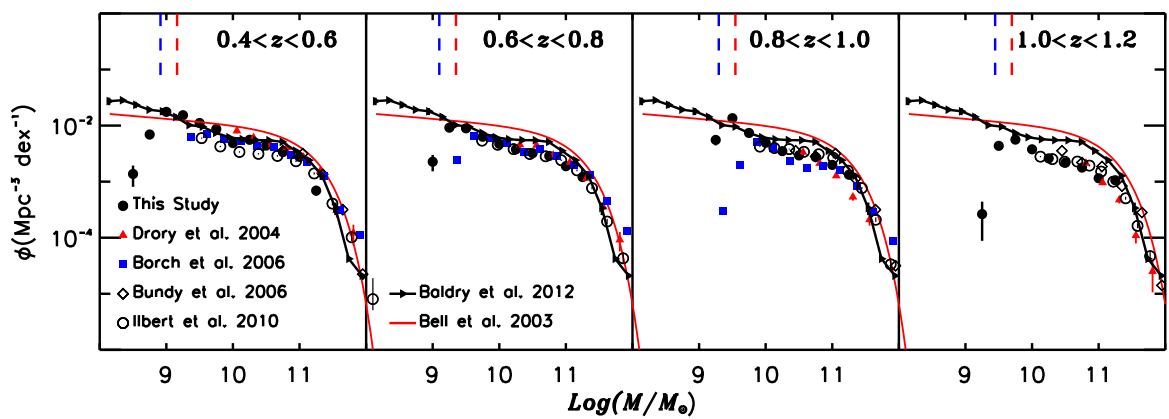}
\caption{Stellar mass functions.  Filled circles show results for the
  3.6\,$\mu$m-selected sample. Other symbols show results for
  comparison samples as indicated in the figure legend. Red solid lines
  indicate the local stellar mass function in \citet{bell2003}, and black solid lines
with arrow are the most updated  local stellar mass function in \citet{baldry2012}. Vertical
  dashed lines indicate the completeness limits for the sample in each
  redshift bin, blue for star-forming galaxies and red for quiescent
  galaxies.  Redshift bins are indicated in each panel.}
\label{f:smf_all}
\end{figure}
\begin{figure}
\plotone{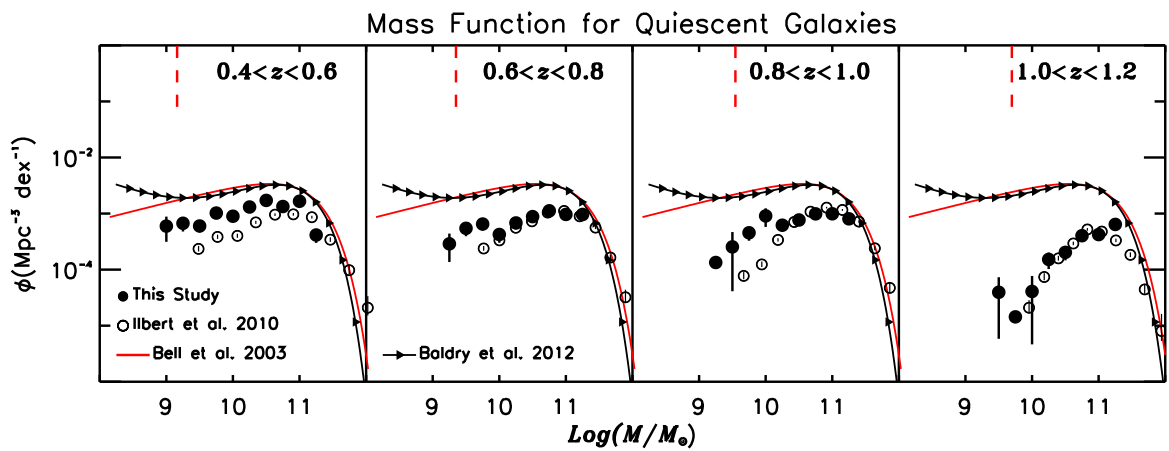}
\caption{Stellar mass functions for quiescent galaxies. Filled
  circles show results for the 3.6\,\micron-selected sample. Vertical
  dashed lines show the completeness limit of the sample. Open
  circles and open diamonds show results for quiescent
  galaxies in the COSMOS field \citep{ilbert2010}.
  Redshift bins are indicated in each panel. Red solid lines show the
  local stellar mass function for quiescent
  galaxiesin \citet{bell2003}, and black solid lines
with arrow are the most updated  local stellar mass function in \citet{baldry2012}.}
\label{f:smf_red}
\end{figure}
\begin{figure}
\plotone{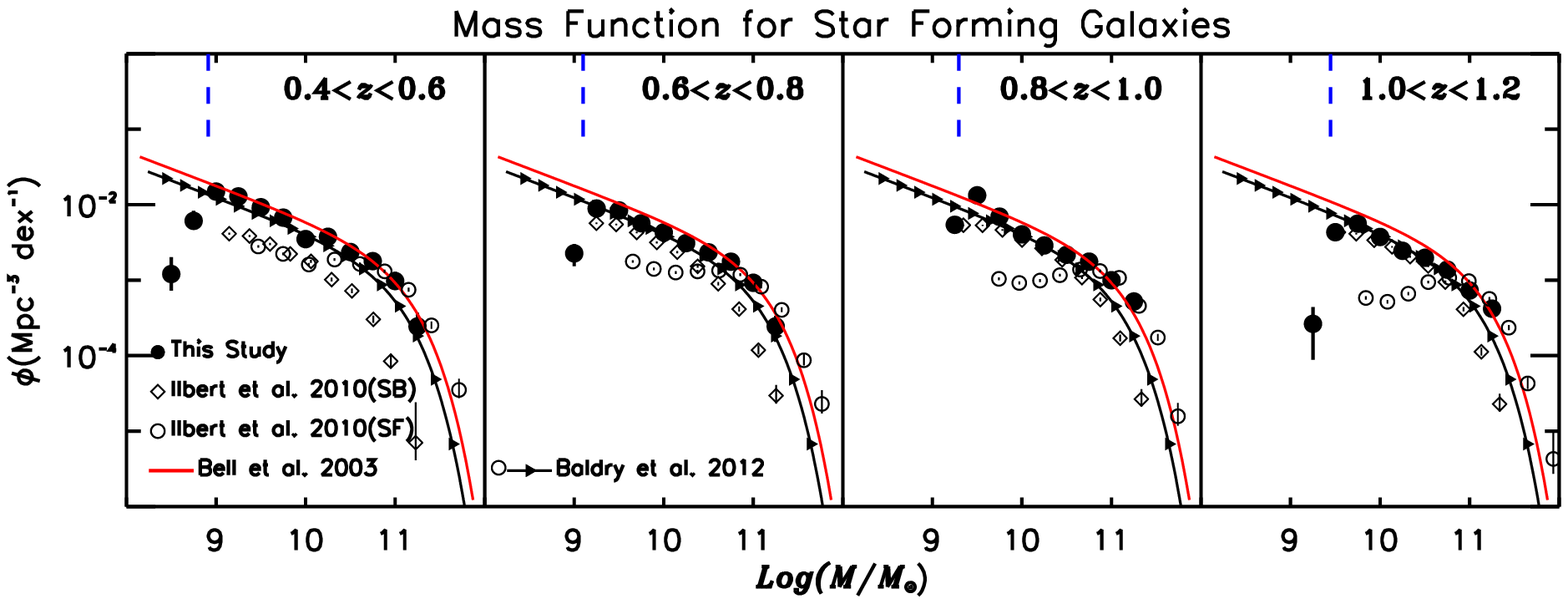}
\caption{Stellar mass functions for star-forming galaxies. Filled
  circles show results for the 3.6\,\micron-selected sample. Vertical
  dashed lines show the completeness limit of the sample. Open
  circles and open diamonds show results for star-forming and
  starburst galaxies in the COSMOS field \citep{ilbert2010}. 
  Redshift bins are indicated in each panel. Red solid lines show the
  local stellar mass function for star-forming
  galaxies\citep{bell2003}, and black solid lines
with arrow are the most updated  local stellar mass function in \citet{baldry2012}.}
\label{f:smf_blue}
\end{figure}

Figures~\ref{f:smf_all}, \ref{f:smf_red}, and \ref{f:smf_blue}
respectively show stellar mass functions for all, quiescent, and
star-forming galaxies.  The mass-to-light ratios for quiescent and
star-forming galaxies differ as shown in Figure~\ref{f:ml} in the
sense that the mass-to-light ratio for quiescent galaxies is roughly
three times larger (Figure~\ref{f:ml}) than for star-forming
galaxies.  The survey mass limit for quiescent galaxies is therefore
correspondingly larger for the fixed 3.6\,\micron\ limiting
magnitude. At the low mass end, star-forming galaxies are the
dominant population in the 3.6\,$\mu$m sample. Thus the mass limit
for the whole sample should be very close to that for star-forming
galaxies. Our stellar mass functions for the 3.6\,$\mu$m selected
sample are consistent with previous estimates \citep{Drory2003,
  borch2006, bundy2006, ilbert2010} but extend below
$10^{9.5}$\,M$_{\odot}$, the lowest stellar mass range so far
explored at these redshifts. Our stellar mass functions for
star-forming galaxies are consistent with those in the COSMOS field
\citep{ilbert2010} except in the bin of $0.4<z<0.6$, where the COSMOS
stellar mass function is lower at $\log(M_*/\Msun)<10.5$.  The mass
functions for quiescent galaxies in the EGS and COSMOS fields are
consistent with each other at the massive end, but our mass functions
for quiescent galaxies are higher at the lower mass end. Our EGS
sample is probably more complete for faint red galaxies because the
IRAC exposure time is nine times longer than that of COSMOS.

\begin{figure}
\plotone{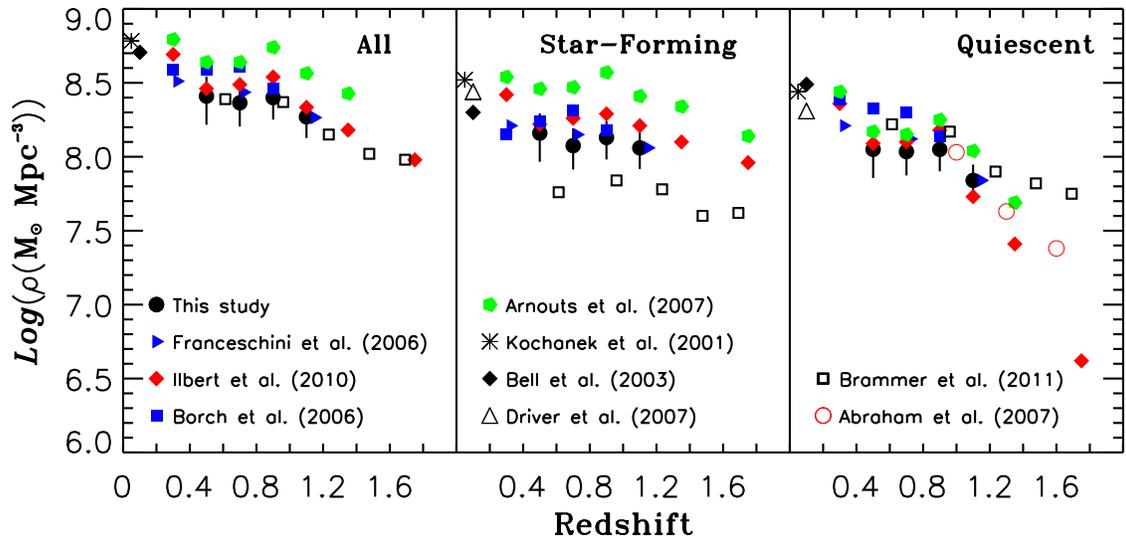}
\caption{Stellar mass density as a  function  of redshift. Black filled
  circles show results for this study, and other symbols show other
  results as indicated in the figure legends.  Panels show total,
  star-forming, and quiescent galaxies respectively, but different
  authors have used different criteria to separate the latter two
  classes.}
\label{f:md_z}
\end{figure}

In comparison with the local stellar mass function \citep{bell2003, baldry2012},
our stellar mass function shows number evolution since $z=1.1$. We
also measured the stellar mass densities for galaxies with
$\log(M_*/\Msun)>9.5$ in each redshift bin 
(Figure~\ref{f:md_z}). The total stellar mass density increased by a
factor of 3--5 from $z=1.1$ to $z\sim0.1$
\citep{kochanek2001,bell2003,driver2006}.
In contrast, our mass function for high-$z$ star-forming galaxies is similar
to local mass functions \citep{bell2003, 
driver2006} (Figure~\ref{f:smf_blue}),  implying very
weak evolution.  The number of star-forming galaxies at the lower mass
end has increased only slightly since $z=1$. The stellar mass
function for quiescent galaxies shows two 
different evolutions in Figure~\ref{f:smf_red}: the absolute number
of massive quiescent galaxies and the slope of the low mass end both
increase at low redshifts. This indicates that massive quiescent
galaxies were still forming at $z<1$, and most low-mass quiescent
galaxies formed much later than their massive counterparts.
 
 The composite stellar mass functions show number evolution since
$z=1.1$ (Figure~\ref{f:smf_all}), indicating continuing mass assembly since that
epoch.  The star-forming galaxy population shows little evolution, and
thus the quiescent population is mainly responsible for the evolution
in the composite mass functions. The quiescent galaxies came from the
star-forming galaxy population, and the amount of mass
assembled in star-forming galaxies is roughly the same as the mass
migrating from the BC to the RS.
Such  mass assembly can also be seen in the stellar mass density
as a function of redshift. Several groups \citep{bell2003,borch2006,
franceschini2006, abraham2007, arnouts2007,ilbert2010} have used stellar
mass densities as a function of redshift to demonstrate how stellar
masses were assembled in galaxies over time
(Figure~\ref{f:md_z}). Our stellar mass densities for both
populations are consistent with the other measurements, and indeed
all surveys yield roughly similar mass densities in the redshift
range of $0.4<z<1.1$ (Figure~\ref{f:md_z}). The mass densities for
star-forming galaxies have large scatter: the VVDS has the highest
\citep{arnouts2007}, and the NEWFIRM has the lowest
\citep{ilbert2010}. All surveys yield similar quiescent galaxy
mass densities within a factor of 2 in this redshift range.  This is
simply due to better photometric redshifts for quiescent galaxies than
star-forming galaxies. The overall mass densities increase by a
factor of $\sim$4 from $z=1.2$ to $z=0$, and the mass densities for
quiescent galaxies also increased by a factor of 5--8 since $z=1.2$.

\section{SUMMARY}
 
A multi-wavelength study of a 3.6\,$\mu$m selected galaxy sample at
$0.4<z<1.2$ with $f_{3.6} >2$\,\mmjy\ in the Extended Groth Strip
region has allowed a new suite of photometric redshifts based on a
neural network method. The distribution of {\large
  $\frac{z_p-z_s}{1+z_s}$} is Gaussian with $\sigma \sim0.025$. Pair
statistics validates the photometric redshifts even for galaxies too
faint to obtain spectroscopic redshifts and shows that the redshifts
have scatter similar to that for the brighter galaxies. The rate of
catastrophic redshift failures is about 3.5\%.

Basic properties measured for the galaxy sample include rest-frame
$U-B$ colors, $B$- and $K$-band absolute magnitudes, and stellar
masses for galaxies in the sample. In the redshift range available,
the IRAC 3.6\,$\mu$m band probes the rest-frame near infrared,
permitting sample selection based on stellar mass and only weakly
biased with respect to quiescent and star-forming galaxies.  Our
sample is complete for galaxies with $\log(M_*/\Msun)>9.5$ at
$0.4<z<1.2$. We divided the sample into quiescent and star-forming
galaxies according to their rest-frame $U-B$ color and
$f_{24}/f_{3.6}$ flux ratio. We then derived $K$-band luminosity
functions and stellar mass functions for quiescent, star-forming, and
all galaxies in the sample at $z=0.5$, 0.7, 0.9, and 1.1.

Our conclusions are:
\begin{enumerate} 
\item
 The $U-B$ vs $B$ color-magnitude diagram cannot by itself
 efficiently separate quiescent and star-forming galaxies. There are
 MIPS 24\,$\mu$m sources in the red sequence in the CM diagram. About
 half of these sources have $f_{24}/f_{3.6} >1$, indicating that they
 are dusty star-forming galaxies. The remaining objects have
 $f_{24}/f_{3.6} <1$, which indicates quiescent galaxies with
 residual star formation or AGN activities.

\item
 The $U-B$--mass diagram shows an absence of quiescent galaxies with
 $\log(M_*/\Msun)<10$ at $1.0<z<1.2$.  The quiescent galaxy mass
 fractions at low masses ($9.5<\log(M_*/\Msun)<10.5$) increase with
 cosmic time from 0\%--10\% at $z=1.1$ to 25\%--35\% at
 $z=0.5$. These quiescent galaxies probably first started to appear at
 $z\sim1.3$.  In comparison, for galaxies an order of magnitude more
 massive 
($10.5<\log(M_*/\Msun)<11$), the quiescent fraction was already 50\%
 of the total stellar mass at $z=1.1$, and this fraction has
 increased only to $\sim$60\% at $z\sim0$.

\item
 Throughout the redshift range $0.4<z<1.2$, $\sim$70\% of the massive
 galaxies ($\log(M_*/\Msun)>11$) are quiescent.  These massive
 quiescent galaxies can have formed in one of two ways: either they
 come directly from star-forming galaxies with similar masses but
 much shorter life times, or they resulted from lower-mass red
 sequence galaxies merging. Some massive galaxies are in a ULIRG
 phase, giving support to the first scenario.

\item
 We derived the $K$-band luminosity functions and stellar mass
 functions in four redshift bins. The cosmic variance in EGS has been
 corrected by comparing the EGS $B$-band luminosity functions with
 the global $B$-band luminosity functions. Our $K$-band luminosity
 functions suggest there may be a luminosity evolution in the
 $K$-band, which needs to be confirmed with larger samples. There is
 number evolution in the stellar mass function, which increases
 from $z=1.1$ to $z=0$. Yet the stellar mass function for
 star-forming galaxies shows very weak evolution compared to strong
 evolution in quiescent galaxies.  This implies that the amount of
 mass assembled in star-forming galaxies is roughly the same as the
 amount of mass in galaxies migrating from the blue cloud to the red
 sequence. The faint end slope for the quiescent galaxy mass function
 has increased since $z=1.1$, indicating that lower mass quiescent
 galaxies formed much later. The mean mass density increased roughly
 a factor of 3 from $z=1.1$ to $z=0$.

\end{enumerate}

\acknowledgements

This work is based in part on observations made with the Spitzer
Space Telescope, which is operated by the Jet Propulsion Laboratory,
California Institute of Technology under a contract with
NASA. Support for this work was provided by NASA through an award
issued by JPL/Caltech. A part of work was done during J.-S. Huang's visit
to Shanghai Normal University supported by the Chinese National Nature Science foundation
Nos. 10878003 and 10833005.

Facilities:
\facility{Spitzer/IRAC}
\facility{Spitzer/MIPS}
\facility{Keck}
\facility{Subaru}



 \clearpage


\clearpage

\clearpage

\clearpage
\clearpage

\clearpage

\clearpage

\clearpage

\clearpage

\clearpage


\clearpage

\clearpage

\clearpage

\clearpage

\clearpage

\clearpage

\clearpage

\clearpage

\clearpage

\clearpage

\clearpage

\clearpage

\clearpage

\clearpage

\begin{table*}[hbt]
{
\begin{center}
\centerline{\sc Table 1}
\vspace{0.1cm}
\centerline{\sc Summary of EGS Optical and Infrared Photometry used for Photo-z Estimation \footnote{For the complete photometry data available in EGS, see \citet{Barro2011}} }
\vspace{0.3cm}
\begin{tabular}{llll}
\hline\hline \noalign{\smallskip}
Band &  Telescope & Depth & Survey\\
\hline
\noalign{\smallskip}
u & MMT &  26.1 & MMT/Megacam EGS Area Survey\citep{Zhao2009}\\
u' & CFHT & 25.7 & Canada-France-Hawaii-Telescope Legacy Survey(CFHTLS)\\
g & MMT & 26.7 & MMT/Megacam EGS Area Survey\citet{Zhao2009}\\
g' & CFHT &  26.5 & Canada-France-Hawaii-Telescope Legacy Survey(CFHTLS)\\
B & CFHT & 25.7 & DEEP2 CFHT Imaging\\
F606W & HST & 26.9  & HST Imaging\\
r' & CFHT & 26.3 &Canada-France-Hawaii-Telescope Legacy Survey(CFHTLS)\\
R & Subaru & 26.1 & Subaru Imaging\citep{Zhao2009}\\
R & CFHT &  25.3 & DEEP2 CFHT Imaging\\
i' & CFHT & 25.9  & Canada-France-Hawaii-Telescope Legacy Survey(CFHTLS)\\
i & MMT & 25.3 & MMT/Megacam EGS Area Survey\citep{Zhao2009}\\
I & CFHT & 24.9 & DEEP2 CFHT Imaging\\
F814W & HST & 26.1 & HST Imaging\\
z' & CFHT & 24.7 & Canada-Hawaii-Telescope Legacy Survey(CFHTLS)\\
z & MMT & 25.3 & MMT/Megacam EGS Area Survey\citep{Zhao2009}\\
J & Palomar & 22.9 & \citep{bundy2006}\\
K$_s$ & Palomar & 22.9  & \citep{bundy2006}\\
K$_s$ & Subaru & 23.7  & \citep{yamada2010}\\
IRAC 3.6\,\micron\ & Spitzer & 23.9  & Spizter GTO+GO \citep{Zheng2012}\\
IRAC 4.5\,\micron\ & Spitzer & 23.9  & Spizter GTO+GO \citep{Zheng2012}\\
IRAC 5.8\,\micron\ & Spitzer & 22.3  & Spizter GTO+GO \citep{Zheng2012}\\
IRAC 8.0\,\micron\ & Sptizer & 22.3  & Spizter GTO+GO \citep{Zheng2012}\\
\noalign{\hrule}
\noalign{\smallskip}
\end{tabular}
\end{center}
}
\label{t:photometry}
\end{table*}
\clearpage

\begin{table*}[hbt]
{
\begin{center}
\centerline{\sc Table 2}
\vspace{0.1cm}
\centerline{\sc EGS $K$-band Luminosity Function}
\vspace{0.3cm}
\begin{tabular}{lllllllll}
\hline\hline \noalign{\smallskip} \multicolumn{1}{c}{}
&\multicolumn{2}{c}{$0.4<z<0.6$} & \multicolumn{2}{c}{$0.6<z<0.8$}
& \multicolumn{2}{c}{$0.8<z<1.0$}&
\multicolumn{2}{c}{$1.0<z<1.2$}
\cr \hline 
$K$ mag & $\log(\phi$) & $\delta\log(\phi$) & $\log(\phi)$ &
$\delta\log(\phi)$ & $\log(\phi)$ & $\delta\log(\phi)$ &$\log(\phi)$
& $\delta\log(\phi)$ \cr

\hline
\noalign{\smallskip}
-18.5 &  -2.67  &  0.30 &       &      &       &      &       &      \cr
-19.0 &  -2.20  &  0.09 & -3.24 & 0.44 &       &      &       &      \cr
-19.5 &  -2.26  &  0.06 & -2.43 & 0.06 & -2.94 & 0.17 &       &      \cr
-20.0 &  -2.29  &  0.07 & -2.45 & 0.05 & -2.42 & 0.05 & -2.97 & 0.18 \cr
-20.5 &  -2.40  &  0.08 & -2.53 & 0.05 & -2.58 & 0.05 & -2.73 & 0.06 \cr
-21.0 &  -2.56  &  0.09 & -2.65 & 0.06 & -2.67 & 0.05 & -2.74 & 0.06 \cr
-21.5 &  -2.68  &  0.10 & -2.73 & 0.06 & -2.81 & 0.06 & -2.82 & 0.06 \cr
-22.0 &  -2.64  &  0.10 & -2.78 & 0.07 & -2.81 & 0.06 & -2.86 & 0.07 \cr
-22.5 &  -2.82  &  0.12 & -2.81 & 0.07 & -2.88 & 0.07 & -2.99 & 0.08 \cr
-23.0 &  -2.79  &  0.12 & -2.94 & 0.08 & -2.98 & 0.07 & -3.02 & 0.08 \cr
-23.5 &  -3.01  &  0.15 & -3.00 & 0.09 & -3.08 & 0.08 & -3.15 & 0.09 \cr
-24.0 &  -3.43  &  0.25 & -3.27 & 0.12 & -3.32 & 0.11 & -3.30 & 0.11 \cr
-24.5 &  -4.23  &  0.62 & -3.91 & 0.25 & -3.84 & 0.20 & -3.90 & 0.22 \cr
-25.0 &         &       & -4.71 & 0.62 & -5.31 & 1.08 & -4.96 & 0.76 \cr
\noalign{\hrule}
\noalign{\smallskip}
\end{tabular}
\end{center}
}
\label{t:klf}
\end{table*}
\clearpage
\begin{table*}[hbt]
{
\begin{center}
\centerline{\sc Table 3}
\vspace{0.1cm}
\centerline{\sc EGS Stellar Mass  Function}
\vspace{0.3cm}
\begin{tabular}{rrrrrrrrr}
\hline\hline \noalign{\smallskip} \multicolumn{1}{c}{}
&\multicolumn{2}{c}{$0.4<z<0.6$} &
\multicolumn{2}{c}{$0.6<z<0.8$} &
\multicolumn{2}{c}{$0.8<z<1.0$}&
\multicolumn{2}{c}{$1.0<z<1.2$}
\cr \hline 

$\log M_*$ & 
$\log( \phi$) &$\delta\log(\phi$) & 
$\log( \phi$) &$\delta\log(\phi$) & 
$\log( \phi$) &$\delta\log(\phi$) & 
$\log( \phi$) &$\delta\log(\phi$) 
\cr \hline \noalign{\smallskip} 

9.00 & -1.81 & 0.11 & -2.65 & 0.22 & & & & \cr 
9.25 & -1.87 & 0.09 & -2.03 & 0.06 & -2.27 & 0.14 & & \cr
9.50 & -2.01 & 0.07 & -2.05 & 0.04 & -1.88 & 0.07 & -2.36 & 0.13 \cr
9.75 & -2.12 & 0.08 & -2.20 & 0.05 & -2.14 & 0.06 & -2.25 & 0.07 \cr
10.00 & -2.37 & 0.10 & -2.33 & 0.06 & -2.32 & 0.07 & -2.42 & 0.07 \cr
10.25 & -2.31 & 0.10 & -2.42 & 0.06 & -2.47 & 0.06 & -2.59 & 0.07 \cr
10.50 & -2.41 & 0.11 & -2.49 & 0.07 & -2.54 & 0.06 & -2.66 & 0.08 \cr
10.75 & -2.52 & 0.13 & -2.54 & 0.07 & -2.57 & 0.07 & -2.75 & 0.08 \cr
11.00 & -2.61 & 0.14 & -2.72 & 0.09 & -2.71 & 0.08 & -2.94 & 0.11 \cr
11.25 & -3.22 & 0.28 & -2.92 & 0.11 & -2.89 & 0.10 & -2.98 & 0.11 \cr

\noalign{\hrule}
\noalign{\smallskip}
\end{tabular}
\end{center}
}
\label{t:mlf}
\end{table*}
\clearpage

\begin{table*}[hbt]
{
\begin{center}
\centerline{\sc Table 4}
\vspace{0.1cm}
\centerline{\sc EGS Stellar Mass  Function for Star-Forming Galaxies}
\vspace{0.3cm}
\begin{tabular}{rrrrrrrrr}
\hline\hline \noalign{\smallskip} \multicolumn{1}{c}{}
&\multicolumn{2}{c}{$0.4<z<0.6$} &
\multicolumn{2}{c}{$0.6<z<0.8$} &
\multicolumn{2}{c}{$0.8<z<1.0$}&
\multicolumn{2}{c}{$1.0<z<1.2$}
\cr \hline 

$\log M_*$ & 
$\log( \phi$) &$\delta\log(\phi$) & 
$\log( \phi$) &$\delta\log(\phi$) & 
$\log( \phi$) &$\delta\log(\phi$) & 
$\log( \phi$) &$\delta\log(\phi$) 
\cr \hline \noalign{\smallskip} 

9.00 &  -3.28  &  0.62 &       &      &       &      &       &      \cr
9.25 &  -3.23  &  0.37 & -3.54 & 0.35 &       &      &       &      \cr
9.50 &  -3.27  &  0.30 & -3.26 & 0.19 & -3.61 & 0.58 & -4.40 & 0.77 \cr
9.75 &  -3.05  &  0.23 & -3.19 & 0.16 & -3.35 & 0.19 & -4.84 & 1.02 \cr
10.00 &  -3.10  &  0.24 & -3.37 & 0.19 & -3.05 & 0.25 & -4.39 & 0.79 \cr
10.25 &  -2.94  &  0.20 & -3.17 & 0.15 & -3.21 & 0.15 & -3.82 & 0.29 \cr
10.50 &  -2.82  &  0.18 & -3.05 & 0.13 & -3.13 & 0.13 & -3.69 & 0.25 \cr
10.75 &  -2.93  &  0.20 & -2.95 & 0.12 & -3.02 & 0.11 & -3.39 & 0.18 \cr
11.00 &  -2.84  &  0.18 & -3.02 & 0.13 & -3.02 & 0.11 & -3.37 & 0.17 \cr
11.25 &  -3.44  &  0.36 & -3.62 & 0.13 & -3.10 & 0.12 & -3.19 & 0.14 \cr

\noalign{\hrule}
\noalign{\smallskip}
\end{tabular}
\end{center}
}
\label{t:mlfq}
\end{table*}
\clearpage

\end{document}